\documentclass[aps,preprint,tightenlines]{revtex4}
\usepackage{graphicx}
\usepackage{subfigure}
\usepackage{pslatex}

\begin{document}


\bigskip

\bigskip

{\large {\bf HETEROGENEOUS MAGNETIC SUPERCONDUCTING SYSTEMS}}

\bigskip

\bigskip

\bigskip

\hspace{2cm} {\bf SERKAN ERDIN}

\hspace{2cm} School of Physics and Astronomy

\hspace{2cm} University of Minnesota, Minneapolis, MN 55455, USA.

\bigskip

\bigskip

\bigskip

\bigskip

\section{Introduction}

  Heterogeneous magnetic superconducting systems (HMSS) represent 
a new class of
nanostructures. They are  made of ferromagnetic (FM) and superconducting
(SC) pieces separated by thin layers of insulating oxides. In contrast
to the case of a homogeneous ferromagnetic superconductor studied during
the last two
decades, the two order parameters, the magnetization and the SC electron
density do not suppress each other \cite{pok1, pl1}.
In HMSS, the interaction between the two order parameters is due to
the magnetic
field created by the magnetic and SC textures. Strong
interaction of the FM and SC systems not only gives rise to a new
class
of novel phenomena and physical effects, but also shows the important
technological promise of devices whose transport properties can be easily
tuned by comparatively weak magnetic fields.

 The interplay between ferromagnetism and superconductivity has
long been the focus of studies both experimental and theoretical
\cite{r1,r2,j1,var}. In 1957, Ginzburg pointed out \cite{ginz} that the
two
phenomena can occur in thin films and wires, due to
the small FM induction  and relatively large SC critical fields.
Afterwards,
Anderson and Suhl \cite{and} described the conditions for both phenomena
to
appear simultaneously in the bulk. The domain-like magnetic structure in
the coexistence phase was noted independently by two different groups
\cite{ful,lar}. Later, Suhl developed the detailed
version of the Landau-Ginzburg theory of ferromagnetic superconductors
\cite{suhl}, e.g. HoMo$_6$S$_8$ and ErRh$_4$B$_4$. The first
experiment in
this context was carried out by Mathias {\it et al.} \cite{math}. The
coexistence of ferromagnetism and superconductivity in the bulk has been
observed recently in the cuprates RuSr$_2$GdCu$_2$O$_{8-\gamma}$ \cite{ru}
and RuSr$_2$Gd$_{1+x}$Ce$_{1-x}$Cu$_2$O$_{10}$ \cite{ru2,ru3} below their
SC transition  temperatures $T_s =  15 - 40$ K and $T_s = 37$ K ( for $x =
0.2$
) respectively.
It is possible to avoid the mutual suppression of the FM and SC order
parameters
by separating them in space \cite{pok1}. Such structures can be made with
modern  nanofabrication techniques. The
proximity effect and spin
diffusion which suppress both order parameters can be easily avoided by
growing a thin insulating oxide layer between the FM and SC components.
Several theoretical studies  have
proposed different possibilities for realization of HMSS: arrays of
magnetic dots on the
top of a SC film \cite{pok1,erdin-array}, single magnetic dot of 
various geometries on top of a SC film \cite{mar,amin1,amin2},  
ferromagnetic/superconducting bilayers (FSB) and multilayers (FSM)
\cite{pok2,ser3,amin3,wei}, magnetic nanorods embedded into a 
superconductor
\cite{pok3,naugle} and semi-infinite HMSS \cite{erdin-semi}.

In HMSS, the strong interaction between FM and SC components stems from
the
magnetic fields generated by the inhomogeneous magnetization and
the supercurrents
as well as SC vortices. SC vortices  were widely studied,
both
experimentally and theoretically, in the literature \cite{vor}. Vortices
in SC films were first studied by J. Pearl \cite{pearl}. He realized that
their current  decays in space  more slowly than those in the bulk.

Similar models of HMSS were  studied theoretically by other groups in
different configurations. Marmorkos {\it et al.} investigated the giant
vortex state created by a magnetic dot, of size on the order of
the coherence
length $\xi$, embedded in a SC film by solving the nonlinear
Landau-Ginzburg
equations \cite{mar}. Kayali \cite{amin1,amin2} and Peeters {\it et al.}
\cite{peeters1} studied vortex creation by both in-plane and out-of-plane 
magnetized 
ferromagnets of different shapes on a SC thin film.  Sasik and Santos {\it 
et al.} considered 
an 
array of FM dots on a SC film \cite{san1,san2}. They treated the dots as
magnetic dipoles, ignoring their real geometry and showed that they excite
and pin SC vortices. Carneiro studied interaction between vortices in 
SC films and magnetic dipole arrays, and showed  that  the arrays with  
antiferromagnetic order is more effective in vortex creation than the 
arrays with ferromagnetic order \cite{carneiro}. Symmetry violation in SC 
thin films with regular arrays of magnetic dots are reported by two 
different studies \cite{erdin-array,fertig}. Bulaevskii {\it et al.} 
discussed the pinning of vortices 
inferromagnet-superconductor multilayers \cite{bul1}. The same group also
examined the effect of the screening magnetic field on a thick magnetic
layer which is placed on top of a bulk superconductor \cite{bul2}. They
found
that the magnetic  domains shrink in the presence of the SC film. In the 
most of theoretical studies, hard magnets are considered. Recently, 
Helseth found  that identical vortices can attract each other in the 
presence of soft magnets \cite{hels-soft}.

To date, only sub-micron magnetic dots
covered by thin SC films have been prepared and studied
\cite{e1,e2,e4,e5,e6,e7,e8}. The experimental samples of FM-SC hybrid
systems were prepared by means of
electron beam lithography and lift-off techniques \cite{ee}. Both in-plane
and
out-of-plane magnetization was experimentally studied. The dots with
magnetization parallel to the plane were  fabricated from
Co, Ni, Fe, Gd-Co and Sm-Co alloys. For the dots with
magnetization
perpendicular to the plane, Co/Pt multilayers were used. The FM dots were
deposited on thin SC films made of either Nb or Pb, whose
transition temperatures are around $7-10$ K. In these experiments, the
effect of commensurability on the transport properties ( e.g.
magnetoresistance oscillations and  matching anomalies ) was observed.
However, this effect is not limited to the magnets interacting with
superconductors and was first found
many years ago by Martinoli and his group \cite{mart}. They studied the
transport properties of SC films with periodically modulated thickness in
an external magnetic field. They found oscillations of critical current
versus magnetic field. Recently, several experimental groups observed
commensurability effects caused by a periodic array of magnetic dots or
holes \cite{met1,met2,met3,met4}.  These results confirm that the FM dots
create
and pin vortices. However, much more interesting and promising would be
effects specific for the HMSS, which are associated with the violation of
the
time reversal symmetry. Some of them include spontaneous currents in the
ground state \cite{pok3,pok5,pok6,pok7}. So far only  one such effect was
experimentally
observed: an asymmetry of the SC hysteresis in the presence of
magnetic
dots \cite{e2}.

In HMSS, the magnetic field 
induced
by  inhomogeneous magnetization penetrates into a superconductor
through SC vortices, while the magnetic field generated by the
supercurrents and SC vortices acts on the magnetic system. The mutual 
interaction between FM and SC subsystems offer interesting physical 
effects such as spontaneous symmetry violation \cite{erdin-array}, 
shrinkage of FM domains and a magnetic domain wall \cite{bul2,ser3, 
helseth}, SC 
transition temperature shift \cite{wei} and Bean-Livingstone-like 
energy barrier for vortices \cite{erdin-semi}. 

We recently developed a method based on London-Maxwell equations to study 
theoretical realizations of HMSS \cite{erdin-common}. In this article, we 
first review this method and its extention to periodic systems and finite 
systems. Next, some of our recent results on a SC film with a FM dot grown 
upon it and domain structures in FSB and further studies in these systems 
are briefly discussed. This review is prepared as a progress report and aims to 
give  reader the basic aspects  of HMSS. This article  is organized 
as 
follows: In the 
next section, a method to calculate inhomogeneous
magnetization and supercurrents including SC vortices in the London
approximation  is presented with great details.  In section II, a 
FM 
dot on top of a SC film and vortex states in the ground state  are 
discussed. Section III is  devoted to the recent theoretical results 
on FSB. We conclude with discussions and summary in section IV. To help 
reader follow easily the theoretical analysis  in 
this 
chapter, we give details of calculations with Bessel functions and series 
in the appendices.

\section{Theory}
In both theoretically proposed and experimentally realized
HMSS,
the magnetic texture interacts with the superconducting (SC) current. An
inhomogeneous
magnetization generates a magnetic field outside the magnets that in
turn generates screening currents in the superconductor, which
subsequently change the magnetic field. The problem  must be
solved self-consistently. In the literature, HMSS have been studied 
through  Landau-Ginzburg equations \cite{mar} and linear London 
method . 
Though the SC vortices are treated more accurately in the former in which 
SC electron density $n_s$ changes in the vicinity of the vortex core, 
solving nonlinear Landau-Ginzburg equations is numerically difficult. In 
London approach, $n_s$ is assumed to be constant, and vortices are treated 
as points. However,  London's approximation is
sufficient
when the sizes of all the
structures in the problem greatly exceed the coherence length
$\xi$, and offers more analytical insight. For this reason, we have been 
studying the several realizations of HMSS through a method based on 
London-Maxwell equations. 
 Here we  present  the  method to calculate
inhomogeneous magnetization and supercurrents including the SC
vortices in the London approximation. In the next section  a  method for the
most general 3-dimensional HMSS is given. In section B, this 
method is applied to
the case of very thin ferromagnetic (FM) and SC films. In section C, the 
method is extended to  periodic heterogeneous magnetic superconducting
systems (PHMSS).
In the last
section, we show how this method can be  applied to semi-infinite and 
finite systems.

\subsection{Three Dimensional Systems}

The total energy of a stationary FM-SC system reads \cite{degennes}
\begin{equation}
U  = \int \bigl [\frac{{\bf B}^2}{8 \pi}  + \frac{m_s n_s {\bf
v}_{s}^{2}}{2} - {\bf B}\cdot {\bf M} \bigr ]dV.\label{en}
\end{equation}
\noindent where ${\bf B}$ is the magnetic induction,  ${\bf M}$ is
the magnetization, $n_s$ is the density of SC electrons, $m_s$ is
their effective mass and ${\bf v}_s$ is their velocity. We  assume
 the SC density $n_s$ and the magnetization ${\bf M}$ to be
separated in space. We also assume that the magnetic field ${\bf
B}$ and its vector-potential  ${\bf A}$  asymptotically approaches
zero at infinity. After the static Maxwell equation ${\bf
\nabla}\times {\bf B} = \frac{4 \pi}{c} {\bf j}$, and ${\bf B} =
{\bf \nabla}\times {\bf A}$ are employed, the magnetic field energy
can be transformed as follows:

\begin{equation}
\int \frac{{\bf B}^2}{8 \pi}  dV=
\int  \frac{{\bf j}\cdot  {\bf A}} {2 c}dV.
\label{en2}
\end{equation}
\noindent Although the vector potential enters explicitly in the
last equation, it is gauge invariant due to current
conservation ${\rm div}{\bf j}=0$. When integrating by parts, we
neglect the surface term. This approximation is correct if the field,
vector potential and current decrease sufficiently fast at
infinity. The current ${\bf j}$ can be represented
as a sum: ${\bf j}={\bf j}_s + {\bf j}_m$ of the SC and magnetic
currents, respectively \cite{landau}:
\begin{equation}
{\bf j}_s  = \frac{n_s\hbar e}{2m_s} \bigl ( \nabla \varphi -
\frac{2\pi}{\phi_0}{\bf A}\bigr ), \label{scurr}
\end{equation}
\begin{equation}
{\bf j}_{m} = c {\bf \nabla} \times {\bf M}.
\label{mcurr}
\end{equation}
\noindent where $\phi_0=hc/2e$  is the SC flux quantum. We separately
consider the  contributions from magnetic and SC currents
to the integral (\ref{en2}), starting with the
integral:
\begin{equation}
\frac{1}{2c}\int {\bf j}_m \cdot {\bf A}dV=\frac{1}{2}\int\bigl(
\nabla\times{\bf M}\bigr ) \cdot{\bf A}dV. \label{magn}
\end{equation}
Integrating by parts and neglecting the surface term again, we
arrive at
\begin{equation}
\frac{1}{2c}\int {\bf j}_m\cdot{\bf A}dV=\frac{1}{2}\int {\bf
M}\cdot{\bf B}dV. \label{magn1}
\end{equation}
We  have omitted the integral over a distant surface:
\begin{equation}
\oint \bigl
({\bf n}\times{\bf M}\bigr ) \cdot{\bf A}dS.
\end{equation}
\noindent  Such an omission is
justified if the magnetization is confined to a limited volume.
But for infinite magnetic systems it may be wrong even in the simplest
problems. 

We next consider the contribution of the SC current
${\bf j}_s$ to the integral (\ref{en2}). In the gauge-invariant
Eq.(\ref{scurr}), $\varphi$  is the phase of the SC carriers
wave-function. Note
that the phase gradient ${\nabla\varphi}$ can be incorporated in
${\bf A}$ as a gauge transformation. The exception is vortex
lines, where ${\varphi}$ is singular. We use the equation
(\ref{scurr}) to express the vector potential ${\bf A}$ in terms of
the supercurrent and the phase gradient:
\begin{equation}
{\bf A}=\frac{\phi_0}{2\pi}\nabla\varphi - \frac{m_sc}{n_se^2}{\bf
j}_s. \label{A}
\end{equation}
Plugging  Eq.(\ref{A}) into  Eq.(\ref{en2}), we find
\begin{equation}
\frac{1}{2c}\int {\bf j}_s\cdot{\bf A}
dV=\frac{\hbar}{4e}\int\nabla\varphi\cdot{\bf j}_s
dV-\frac{m_s}{2n_se^2}\int j_s^2dV. \label{ensup}
\end{equation}
Since the superconducting current is
\begin{equation}
{\bf j}_s=en_s{\bf v}_s.
\end{equation}
\noindent The last
term in Eq.(\ref{ensup})
equals the negative of the  kinetic energy and thus exactly compensates
the
kinetic energy in the initial expression for the energy
(\ref{en}). Collecting all the remaining terms, we obtain the
following expression for the total energy:
\begin{equation}
U  = \int \bigl[\frac{n_s\hbar^2}{8m_s}({\nabla\varphi})^2 -
\frac{ n_s\hbar e}{4m_sc}{\nabla\varphi} \cdot{\bf A} - \frac{{\bf
B}\cdot {\bf M}}{2}\bigr] dV. \label{en3}
\end{equation}
This expression is complete except for a possible surface
term for infinite magnetic systems. Note that integration volume  includes
both superconductors and  magnets. Eq. (\ref{en3})
allows us to separate the energy of the vortices, the energy of
magnetization and the energy of their interaction. Indeed, as we
noted earlier, the phase gradient can be ascribed to the
contribution of vortex lines alone. It can be represented as a sum
of independent integrals over distinct vortex lines. The
vector-potential and the magnetic field can also be presented as a
sum of magnetization induced and vortex induced parts: ${\bf
A}={\bf A}_{m}+{\bf A}_{v}$, ${\bf B}={\bf B}_{m}+{\bf B}_{v}$,
where ${\bf A}_{k}$, ${\bf B}_{k}$ (the index $k$ is either $m$ or
$v$) are determined as solutions of the London-Maxwell equations
generated by the magnetization and the vortices.
The effect of the SC screening of the magnetic field due to
the magnetization is already included in the vector fields ${\bf
A}_{m}$ and ${\bf B}_{m}$. If such separation of fields is applied, then
the total energy, (\ref{en3}) becomes a sum of terms
containing  vortex contributions alone,  magnetic contributions
alone and the interaction terms. The purely magnetic component can
be represented as a non-local quadratic form of the magnetization.
The purely superconducting part becomes a non-local double
integral over the vortex lines. Finally, the interaction term may
be presented as a double integral over the vortex lines and the
volume occupied by the magnetization, and is bi-linear in
magnetization and vorticity.

\subsection{Two Dimensional Textures and Vortices}
Below we show a detailed analysis in the case of parallel FM and
SC films, both very thin and positioned close to each other.
Neglecting their thickness, we assume both films to be located
approximately at $z = 0$. In some cases we need a higher degree of
accuracy. We then introduce a small distance $d$ between the films,
which in the end approaches zero. Although the thickness of
each film is assumed to be small, the 2-dimensional densities of
super-carriers $n_s^{(2)}=n_sd_s$ and magnetization ${\bf m}={\bf
M}d_m$ remain finite. Here $d_s$ is the thickness of the SC film
and $d_m$ is the thickness of the FM film. The 3d super-carrier
density in the SC film is $n_s({\bf R}) = \delta (z)
{n}^{(2)}_s({\bf r})$ and the 3d magnetization in the FM film is
${\bf M}({\bf R}) = {\delta(z-d)}{\bf m}({\bf r})$, where ${\bf
r}$ is the two-dimensional radius-vector and the $z$-direction is
chosen to be perpendicular to the films. In what follows the 2d SC
density ${n}^{(2)}_s$ is assumed to be a constant and the index
(2) is omitted.  The energy (\ref{en3}) for this special case
takes the following form:

\begin{equation}
U  = \int \bigl[\frac{n_s\hbar^2}{8m_s}({\nabla\varphi})^2 -
\frac{ n_s\hbar e}{4m_sc}{\nabla\varphi} \cdot{\bf a} - \frac{{\bf
b}\cdot {\bf m}}{2}\bigr] d^2 {\bf r},
\label{en4}
\end{equation}
where ${\bf a} = {\bf A}({\bf r},z = 0)$ and ${\bf b} = {\bf
B}({\bf r},z = 0)$. The vector potential satisfies the Maxwell-London
equation, which is derived from the static Maxwell equation ${\bf \nabla}
\times {\bf B} = \frac{4 \pi}{c} {\bf j}$, where  ${\bf j}$ is  the
total current
density on the surface of the superconductor, and is given by ${\bf j} =
({\bf j}_s + {\bf j}_m)\delta(z)$. The supercurrent and the magnetic
current densities are given in Eqs.(\ref{scurr}, \ref{mcurr}). Using
${\bf B} =
{\bf \nabla}\times{\bf A}$, the Maxwell-London equation reads
\begin{equation}
\label{vec}
\nabla\times(\nabla\times {\bf A}) = - \frac{1}{\lambda} {\bf A}
\delta (z) + \frac{ 2\pi\hbar  n_se}{m_s
c}{\nabla\varphi}\delta(z)
 + 4\pi\nabla\times ({\bf m}\delta(z)).
\end{equation}
\noindent Here $\lambda = \lambda_L^2/d_s$ is the effective
screening length for the SC film, and  $\lambda_L = (\frac{m_s c^2}{4 \pi
n_s
e^2})^\frac{1}{2}$ is the London
penetration depth \cite{abr}.

According to our general arguments, the term proportional to
${\nabla\varphi}$ in Eq. (\ref{vec}) describes vortices.  A plane
vortex characterized by its vorticity $n$ and by the position  ${\bf
r}_0$ of its
center on the plane, contributes a singular term to
${\nabla\varphi}$:

\begin{equation}
\nabla\varphi_0({\bf r,r}_0)=n\frac{\hat z\times ({\bf r}-{\bf
r}_0) } {\vert {\bf r}-{\bf r}_0 \vert^2}, \label{vortphase}
\end{equation}

\noindent and generates a Pearl vortex vector potential(see Appendix A 
for details):
\begin{equation}
{\bf A}_{v0}({\bf r}-{\bf r}_0,z)
=
\frac{n \phi_0}{2 \pi}
\frac{\hat z\times ({\bf r}-{\bf r}_0) }{\vert {\bf r}-{\bf r}_0
\vert}
\int_{0}^{\infty} \frac{J_1 ( q\vert{\bf r} - {\bf
r}_0\vert) e^{-q| z |}} {1 + 2\lambda q}dq, \label{vec2}
\end{equation}
where $J_1 ( x )$ is Bessel function of the first order. Different
vortices contribute independently in the
vector potential and magnetic field. In the limit of zero film
thickness, the usual Coulomb
gauge, ${\rm div}{\bf A}=0$, leads to
strong singularity in the vector potential, due to the surface currents
in the SC and FM films, which leads to the discontinuity at $z=0$.
Therefore, we choose to employ another
gauge $A_z = 0$. The calculations
become simple in the Fourier-representation. Following the
previous section, we write the
Fourier transform of the vector potential ${\bf A_k}$ as a sum
${\bf A_k}={\bf A}_{m{\bf k}}+{\bf A}_{v{\bf k}}$ of independent
contributions from magnetization and vortices. Using the following
definitions of the  Fourier transform:
\begin{eqnarray}
{\bf A}_{\bf k} &=& \int {\bf A} ({\bf r},z) e^{- i {\bf q} \cdot {\bf
r}- i k_z z} d^3 r, \\
{\bf a}_{\bf q} &=& \int {\bf a} ({\bf r},z=0) e^{- i {\bf q} \cdot {\bf
r}} d^2 r.
\end{eqnarray}
The equation for the
magnetic part of the vector-potential reads
\begin{equation}
  {\bf k}({\bf k\cdot A}_{m{\bf k}})-k^2{\bf A}_{m{\bf k}}=\frac{{\bf
  a}_{m{\bf q}}}{\lambda}-4\pi i{\bf k}\times{\bf m_q}e^{ik_zd},
  \label{A-mag}
\end{equation}
where ${\bf q}$ is the projection of the wave vector ${\bf k}$ onto
the plane of the films: ${\bf k}=k_z{\hat z}+{\bf q}$. An
arbitrary vector field ${\bf V_k}$ in wave-vector space can be
fixed by its coordinates in a local frame of reference formed by
the vectors ${\hat z},{\hat q},{\hat z}\times{\hat q}$:
\begin{equation}\label{local}
  {\bf V_k}= V_{\bf k}^z{\hat
z}+ V_{\bf k}^{\parallel}{\hat q} + V_{\bf k}^{\perp}({\hat
z}\times{\hat q}).
\end{equation}
The solution to equation (\ref{A-mag}) with $A_z = 0$ is found  by taking 
the
inner product of equation (\ref{local}) with $\hat q$, $\hat z$ and $\hat
z \times \hat q$, respectively as below:

\begin{equation}\label{A-par2}
  A_{m{\bf k}}^{\parallel}=-\frac{4\pi im_{\bf
  q}^{\perp}}{k_z}e^{ik_zd} - \frac{a_{m{\bf q}}^{\parallel}}{k_z^2
\lambda},
\end{equation}
\begin{equation}\label{A-par}
  A_{m{\bf k}}^{\parallel}=-\frac{4\pi im_{\bf
  q}^{\perp}}{k_z}e^{ik_zd},
\end{equation}
\begin{equation}
 A_{m{\bf k}}^{\perp}=-\frac{1}{\lambda k^2}a_{\bf
 q}^{\perp}+\frac{4\pi i\left( k_zm_{\bf q}^{\parallel}-qm_{{\bf
 q}z}\right)}{k^2}e^{ik_zd}.
\label{A-perp}
\end{equation}
Integration of the latter equation over $k_z$ gives the
perpendicular component of ${\bf a_{\bf q}}^{(m)}$:
\begin{equation}
  a_{m{\bf q}}^{\perp}=-\frac{4\pi\lambda q(m_{\bf q}^{\parallel}+im_{{\bf
  q}z})}{1+2\lambda q}e^{-qd}.
  \label{a-perp}
\end{equation}
It follows from Eqs.(\ref{A-par2}, \ref{A-par}) that $a_{m{\bf
q}}^{\parallel}=0$. Note that  Eq. (\ref{A-par}) for the parallel
component of the vector potential $A_{m{\bf k}}^{\parallel}$ does
not contain any information on the SC film. This component
corresponds to zero magnetic field outside the FM film.
Therefore, it is not essential for our problem.
The vortex part of the vector potential ${\bf A}_{v{\bf k}}$ also
has not a   $z$-component since the supercurrents flow in the
plane. The vortex-induced vector potential is \cite{abr}
\begin{equation}
  {\bf A}_{v{\bf k}}=\frac{2i\phi_0(\hat{q}\times\hat{z})F({\bf q})}{{\bf
k}^2(1+2\lambda q)},
  \label{A-vortex}
\end{equation}
\noindent where $F({\bf q})=\sum_{j} n_j e^{i{\bf q}\cdot{\bf r}_j}$ is
the vortex
form-factor; the index $j$ labels the vortices, $n_j$ denotes the
vorticity of the $j$th vortex  and ${\bf r}_j$ are
coordinates of the vortex centers. The Fourier-transform of the
vortex-induced vector potential at the surface of the SC film
${\bf a}_{v{\bf q}}$ reads
\begin{equation}
 {\bf a}_{v{\bf q}}=\frac{i\phi_0(\hat{q}\times\hat{z})F({\bf
q})}{q(1+2\lambda
 q)}.
 \label{a-vortex}
\end{equation}
We express the energy (\ref{en4}) in terms of the fields and
vector-potential Fourier-transforms separating the purely
magnetic, purely vortex and the interaction parts:
\begin{equation}
  U=U_{vv}+U_{mm}+U_{mv}.
  \label{en5}
\end{equation}
The vortex energy $U_{vv}$ is the same as it would be in the absence
of the FM film:
\begin{equation}
U_{vv}=\frac{n_s\hbar^2}{8m_s}\int \nabla\varphi_{-{\bf
q}}\cdot(\nabla\varphi_{\bf q}-\frac{2\pi}{\phi_0}{\bf a}_{v{\bf
q}})\frac{d^2q}{(2\pi)^2} \label{en-v}
\end{equation}
However, the magnetic energy $U_{mm}$:
\begin{equation}
 U_{mm}=-\frac{1}{2}\int{\bf m_{-\bf q}}\cdot{\bf b}_{m{\bf q}} \frac{d^2
q}{2
\pi^2}
  \label{en-m}
\end{equation}
contains the screened magnetic field ${\bf b}$ and therefore
differs from its value in the absence of the SC film, but it does
not depend on the vortex positions. The interaction energy reads
\begin{eqnarray}
  U_{mv}&=&\nonumber
-\frac{n_s\hbar e}{4m_sc}\int (\nabla\varphi)_{-{\bf
  q}}\cdot{\bf a}_{m{\bf q}}\frac{d^2q}{(2\pi)^2}
\\&-&
\frac{1}{2}\int{\bf
  m}_{-{\bf q}}\cdot{\bf b}_{v{\bf q}}\frac{d^2q}{(2\pi)^2}.
  \label{en-mv}
\end{eqnarray}
Note that only the form-factor $F({\bf q})$ conveys any information
about the vortex arrangement.

\subsection{Periodic Systems}

A periodic
heterogeneous magnetic superconducting system (PHMSS) such as a magnetic 
dot
array or a periodic domains in ferromagnet-superconductor bilayers can
be studied with the
method described in the previous section. However, it is necessary to
modify
the equations
given above for the periodic structures of interest. In this section, we
extend the above  method to study PHMSSs. In
doing so, we still assume that PHMSSs are made of very thin magnetic
textures with the magnetization perpendicular to the plane  and SC films.
Their energy is calculated over the surface of
the SC film. We start with the energy of
the 2d systems (\ref{en4}).
In the plane of the PHMSS, the magnetic
field ${\bf b}$,
the magnetization ${\bf m}$, the
phase gradient ${\bf \nabla} \varphi$ and the vector potential ${\bf a}$
are 2d periodic
functions.
Therefore, we need to express  them in terms of   Fourier series.
For any function $f({\bf r})$, the  Fourier expansion is given by
\begin{equation}
{\bf f} ({\bf r}) = \sum_{\bf G} {\bf f}_{\bf G} e^{i {\bf
G}
\cdot {\bf r} } \hspace{1cm} {\bf f}_{\bf G} = \frac{1}{\cal A} \int {\bf
f} ({\bf r}) e^{-i {\bf
G}
\cdot {\bf r} } d^2 {\bf r}.
\label{discfo}
\end{equation}
\noindent The ${\bf G}$ s
are  the reciprocal vectors of the
periodic
structure of interest, and ${\cal A}$ is the elementary cell area. We
first express ${\bf a}$, ${\bf b}$, ${\bf
m}$
and ${\bf \nabla} \varphi$ in terms of the Fourier series as in
(\ref{discfo}), then substitute them back in  (\ref{en4}). Performing the
integral over the infinite area of the  2d system and  using
$\int e^{i
({\bf G} +{\bf
G}^\prime ) \cdot {\bf r}} d^2 r = {\cal A} \delta_{{\bf G},-{\bf
G}^\prime}$, we obtain  the energy per unit cell $u$ for the 2d   PHMSS
expressed in terms
of
 Fourier components as follows:

\begin{equation}
u = \sum_{\bf G} [\frac{n_s \hbar^2}{8 m_s} |(\nabla
\varphi)_{\bf G}|^2 -
\frac{n_s \hbar e}{4 m_s c} ({\bf \nabla} \varphi)_{\bf
G}\cdot {\bf
a}_{-\bf G} - \frac{ {\bf b}_{\bf G} \cdot {\bf m}_{-{\bf
G}}}{2} ].
\label{enrealsum}
\end{equation}

\noindent The   
Fourier coefficients of the vector potentials, ${\bf a}_{m}$ and
${\bf
a}_{v}$
 for both the magnetic part and
the vortices in terms of continuous Fourier vectors are already given in
(\ref{a-perp}) and (\ref{a-vortex}). They can be rewritten in terms
of
reciprocal vectors as

\begin{equation}
{\bf a}_{m {\bf G}} = \frac{4 \pi \lambda i {\bf G} \times \hat z
m_{z\bf
G}}{1 + 2 \lambda G}, \label{am-G}
\end{equation}
\begin{equation}
{\bf a}_{v{\bf G}} = \frac{i\phi_0(\hat{G}\times\hat{z})F_{\bf
G}}{{\cal A}G(1+2\lambda
 G)}.
 \label{av-G}
\end{equation}  

\noindent Uusing  the
Fourier coefficients of
the
magnetic field and the phase gradient:
\begin{equation}
{\bf b}_{\bf G} = i {\bf G} \times {\bf a}_{\bf G},
\label{bG}
\end{equation}
\begin{equation}
({\bf \nabla} \varphi)_{\bf
G} = 2 \pi i ({\bf G} \times \hat z F_{\bf
G})/ ({\cal A} G^2),   
\label{phase}
\end{equation}
\noindent  and replacing the vector potentials in
(\ref{enrealsum}) by (\ref{am-G}) and (\ref{av-G}), the energy of the
PHMSS per unit cell
is
found
term by
term as
\begin{eqnarray}
u_{vv} &=& \frac{\phi_0^2}{4 \pi{\cal A}^2} \sum_{\bf G} \frac{
|F_{\bf
G}|^2}{G ( 1 + 2 \lambda G)}, \label{hv} \\
u_{mv} &=& - \frac{\phi_0}{\cal A} \sum_{\bf G} \frac{m_{z {\bf
G}}F_{-\bf G}}{1 + 2 \lambda G}, \label{hmv}\\
u_{mm} &=& -2 \pi \lambda \sum_{\bf G} \frac{G^2 |{\bf m}_{z\bf G}|^2}{1 +
2
\lambda G}.
\label{hmm}
\end{eqnarray}

\subsection{Finite Systems}

In
the most of
theoretical
studies, SC subsystem is considered to be
infinite size for the sake of computational simplicity. Although it is 
relatively hard to handle SC system's boundaries in both 
Landau-Ginzburg and London equations, several  groups  have studied finite 
or semi-finite realizations of HMSS.  
We recently considered semi-infinite HMSS  
elsewhere \cite{erdin-semi}, by benefiting from the ideas of 
Kogan's study of Pearl vortex near the edge of SC film\cite{kogan}. 
Here, we propose an alternative method to treat semi-infinite and finite 
realizations of HMSS. To this end, we modify London-Maxwell equation (see 
Eq.(\ref{vec})). The boundary 
condition can be incorporated into Maxwell-London equation by using 
step function, namely supercurrent is  zero outside the SC 
system's boundary. Together with step function, Maxwell-London 
equation reads
\begin{equation}
{\bf \nabla} \times {\bf \nabla} \times {\bf A} = \frac{4 \pi}{c} ( {\bf 
j}_s \theta({\bf r}^\prime - {\bf r} ) + {\bf j}_m ), 
\label{finiteeq}
\end{equation}
where ${\bf j}_s = \frac{c \Phi_0}{8 \pi^2 \lambda} {\bf \nabla} 
\varphi  - \frac{c}{4 \pi \lambda} {\bf a} )$ and 
$\theta({\bf r}^\prime - {\bf r} )$ is step function that equals 0 or 1 
when ${\bf r}$ is 
greater than the 
boundary's position ${\bf r}^\prime$ or otherwise. For semi-infinite 
system in which 2-d SC film lies on x-y  
plane and its edge is located at x=0, step function is $\theta(x)$. For a 
finite circular SC disk of radius R, step function becomes $\theta(R-r)$. 
Eq.(\ref{finiteeq}) can be solved by similar techniques used in the 
previous sections, however its solution gives rather complicated integral 
equations. We leave the details of the solution to further works.

\section{FM Dot on SC Film}
\begin{figure}[htb]
\begin{center}
\includegraphics[angle=0,width=3.5in,totalheight=3.5in]{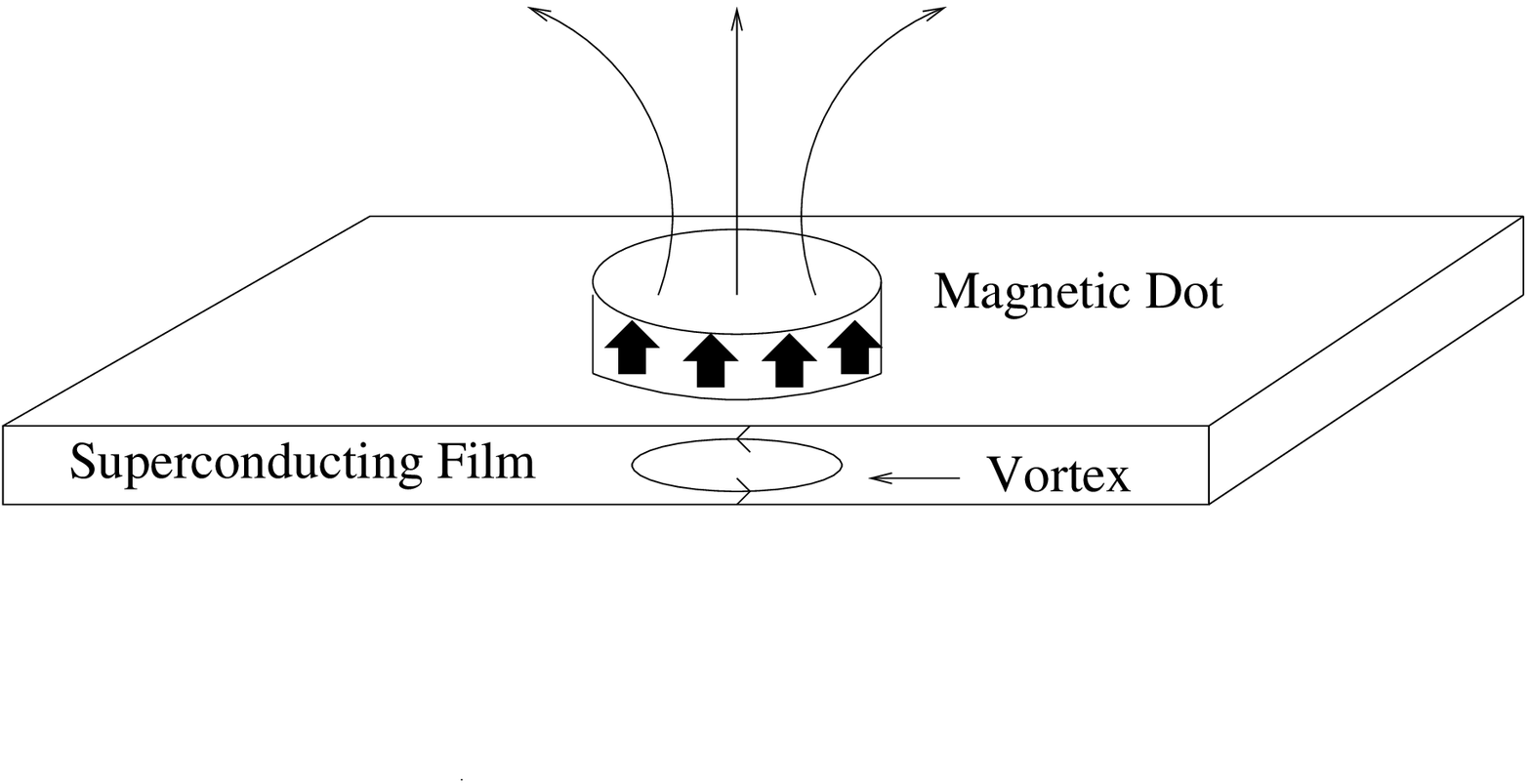}
\caption{\label{fig:mdot}Magnetic dot on a superconducting film. }
\end{center}
\end{figure}

In this section, we review the studies related to  the ground state of a 
superconducting 
(SC)
film in the presence of a circular FM dot grown upon it (see (see Fig. 
\ref{fig:mdot})). In this case,  the magnetization is
assumed to be fixed and  homogeneous within  the dot, and directed
perpendicular to the SC film. This problem is previously discussed 
elsewhere
\cite{erdin-common}, in which we predicted the geometrical pattern formed
by vortices in the ground state.
Here, we give further details of analysis and some
new results. The problems we
discuss are: i) under what conditions do vortices 
appear in
the
ground
state; ii) where they appear, and iii) the magnetic fields
and currents in these states. As in the previous section, we
assume the SC film to be a very thin plane, and infinite in the
lateral directions. Since the magnetization is confined within the
finite dot, no integrals over infinitely remote surfaces or
contours arise. In the next section, we treat the first case using  
the
method described in
the previous chapter. 

Let both SC and FM films be infinitely thin and placed at
heights $z=0$ and $z=d$, respectively. The SC film is infinite in
lateral directions, whereas the FM film is finite and has the shape 
of
circle with  radius $R$ (magnetic dot). The 2d
magnetization of the magnetic dot is ${\bf m(r)} = m \hat{z}\sigma
(R-r)\delta(z-d)$, where $\sigma(x)$ is the step function, equal to
$+1$ for $x>0$ and $0$ for $x<0$. We first find the
vector potential and magnetic field induced by the dot in the
presence of the SC film, using Eqs.
(\ref{A-perp},\ref{a-perp}). The Fourier-component ofd at
magnetization necessary for this calculation is
\begin{equation}
  {\bf m_k}=\hat{z}\frac{2\pi mR}{q}J_1(qR)e^{ik_zd},
 \label{mperp-Fourier}
\end{equation}
where $J_1(x)$ is the Bessel function. From (\ref{a-vortex}) with
(\ref{mperp-Fourier}),
the  calculations employ
Fourier-transform of
the vector potential at the superconductor surface:
\begin{equation}
  a_{m{\bf q}}^{\perp}=-\frac{i8\pi^2\lambda
  mR}{1+2\lambda q}J_1(qR).
\label{aperp}
\end{equation}
In the last equation we have replaced $e^{-qd}$ by 1. The
Fourier-transform of
the vector potential reads
\begin{equation}
  A_{m{\bf k}}^{\perp}=\nonumber
-\frac{i8\pi^2 mRJ_1(qR)}{k^2}
 \left[ e^{-qd}\frac{2q\lambda}{1+2\lambda q}+(e^{ik_zd}-e^{-qd})
  \right].
 \label{Aperp}
\end{equation}

Though the difference in the round brackets in equation
(\ref{Aperp}) appears to be small (recall that $d$ must be
set to  zero in the final answer), we cannot neglect it since it
implies a finite, not small discontinuity in the parallel
component of magnetic field at the two film faces. From equation
(\ref{Aperp}) we can immediately find the Fourier-transforms of the
magnetic field components via
\begin{equation}
  B_{m{\bf k}}^z=iqA_{m{\bf k}}^{\perp};\,\,\,B_{m{\bf
k}}^{\perp}=-ik_zA_{m{\bf
  k}}^{\perp}.
  \label{Bperp}
\end{equation}
The inverse Fourier-transformation of Eqs.
(\ref{Bperp},\ref{Aperp}) gives the magnetic field in real space:
\begin{equation}
B_m^z({\bf r},z)=4\pi\lambda mR
\int_0^{\infty}\frac{J_1(qR)J_0(qr)e^{-q|z|}}{1+2\lambda q}q^2dq,
\label{Bz-coord}
\end{equation}
\begin{equation}
\label{Br-coord}
B_m^r({\bf r},z) =  -2\pi mR
 \int_0^{\infty}J_1(qR)J_1(qr)e^{-q|z|}
 \left[\frac{2q\lambda
 }{1+2q\lambda}{\mathrm{sign}}(z)+{\mathrm{sign}}(z-d) -
{\mathrm{sign}}(z)\right]qdq,
\end{equation}
where ${\mathrm{sign}}(z)$ is the function equal to the sign of its
argument. Note that $B_m^r$ has discontinuities at $z=0$ and $z=d$
due to surface currents in the SC and FM films respectively;
whereas, the normal component $B_m^z$ is continuous.

Symmetry arguments imply that a vortex, if it appears, must be
located at the center of the dot. Indeed, for $R \gg\lambda$, an
analytical calculation shows that the central position of the
vortex provides minimal energy. We have checked numerically that
the central position is always energy favorable for one vortex. This
fact is not trivial since the magnetic field of the dot is
stronger near its boundary and a violation of symmetry could be
naively expected. However, the gain of energy due to interaction
of the magnetic field generated by the vortex with the magnetization 
of
the dot
decreases when the vortex approaches the boundary.

\begin{figure}[htb]
\begin{center}
\includegraphics[angle=270,width=3.5in,totalheight=3.5in]{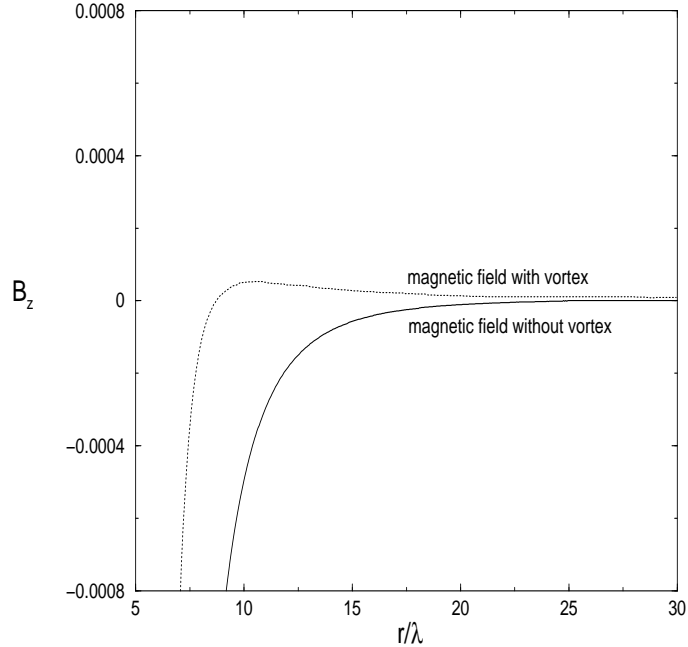}
\caption{\label{fig:vor}Magnetic field of a dot with and without
vortex for $R/\lambda = 5$ and $\phi_0/8 \pi^2 m R = 0.05$. }
\end{center}
\end{figure}

Another interesting problem is the sign of the perpendicular
component of the magnetic field. The vector potential generated by a
vortex is given by Eq. (\ref{A-vortex}) with $F({\bf q})=1$.  The
perpendicular
component of the magnetic field generated by the vortex is
\begin{equation}
\label{Bzv-coord}
B_v^z=\frac{\phi_0}{2\pi}\int_0^{\infty}\frac{J_0(qr)e^{-q|z|}}{1+2\lambda
  q}qdq.
\end{equation}
Numerical calculation based on Eqs. (\ref{Bz-coord},
\ref{Bzv-coord}) shows that, in the presence of the vortex
centered at $r=0$, $B_z$ on the SC film $(z=0)$ changes sign for
some $r>R$ (see Fig. \ref{fig:vor}), but it is negative for all
 $r > R $ without the vortex.
The physical explanation of this fact is as follows. The dot
itself is an ensemble of parallel magnetic dipoles. Each dipole
generates a magnetic field  whose $z$-component on the plane passing
through the dot has  sign opposite to the dipolar moment.
However, the field exactly over and under the dipole has the same
sign as the dipole and is strongly singular. The fields from
different dipoles compete at $r<R$, but they have the same sign
at $r>R$. The SC current tends to screen the  magnetic field of
the magnetic dot  and have the opposite sign. The field generated by 
a
vortex at
large
distances decays more slowly than the screened dipolar field
($1/r^3$ vs. $1/r^5$ ). Thus, the sign of $B_z$ is opposite to the
magnetization at small values of $r$ (but larger than $R$) and
positive at large $r$. Measurement of the magnetic field near the
film may serve as a diagnostic tool to detect a SC vortex bound by
the dot. To our knowledge, so far there are no experimental
measurements of this effect.

The energy of the system in the presence of many vortices with
arbitrary
vorticities $n_i$ can be
calculated using Eqs.(\ref{en5}-\ref{en-mv}). The appearance of N
vortices with arbitrary positions ${\bf r}_i$ in the system changes
the energy by an amount:
\begin{equation}\label{change}
  \Delta_N = \sum_{i=1}^{N} n_i^2 \varepsilon_v +
\frac{1}{2} \sum_{i\neq j}^N n_i n_j \varepsilon_{vv} (r_{ij}) +
\sum_{i=1}^N
n_i \varepsilon_{mv}( r_i).
\end{equation}
Here $\varepsilon_v = \varepsilon_0\ln (\lambda /\xi)$ is the
energy of a vortex without a magnetic dot, $\varepsilon_0 =
{\phi^2_0 /(16 \pi^2 \lambda)}$ and  $\varepsilon_{vv}$ is the
vortex-vortex interaction and $\varepsilon_{mv}$ is the vortex-magnetic 
dot 
interaction.
Substituting (\ref{a-vortex}) into the vortex energy $\varepsilon_{vv}$ of 
(\ref{en-v}), we get

\begin{equation}
\varepsilon_{vv} (r_{ij}) = \frac{\varepsilon_0}{\pi} \left[H_0
(\frac{r_{ij}}{2
\lambda}) - Y_0 (\frac{r_{ij}}{2  \lambda})\right ],
\end{equation}
\noindent where $r_{ij} = |{\bf r}_j - {\bf r}_i|$ , $H_0 ( x )$ and $Y_0 
( x )$ are the Struve function of the zeroth order and the modified 
Bessel function 
of the second kind of the
zeroth order, respectively \cite{abrom}.  For $\varepsilon_{mv}$ of
(\ref{en-mv}), direct substitution of the
vector potential, magnetic field and the phase gradient (see Eqs.
(\ref{aperp},\ref{Bz-coord})) into
(\ref{en-mv}) gives
\begin{equation}\label{epsilon-mv}
  \varepsilon_{mv} (r_i) =
  -m\phi_0 R\int_0^{\infty}\frac{J_1(qR) J_0(qr_i)dq}{1+2\lambda q}.
\end{equation}

 In order for  N vortices to appear, the necessary
condition is that $\Delta_N < 0$ and $\Delta_N < \Delta_{N-1}$. Using
this criteria, we can determine in what configurations and order the
vortices appear. To this end, we study only  vortices with positive
vorticity which are situated under the dot. Under the assumption that the
dot's diameter  is  larger than $\xi$, it is reasonable to
think that
vortices with  multiple
vorticities, the  so-called giant vortices, do not appear,
since the vortex  energy grows as the square of its vorticity (see
Eq.(\ref{change})). For large dots $R>>\lambda$ and sufficiently small
$m$,
the giant vortex is definitely energy unfavorable. However  for $R <
\lambda$
and $m R >> \phi_0$, it can be favorable. This question has not yet
been analyzed
completely.

The next step is to minimize (\ref{change}) with respect to the positions
of the vortices. We first start with one vortex.
It turns out that it  appears at the center of the dot. $\Delta_1$
is a function of two dimensionless
variables   $m\phi_0/\varepsilon_v$ and $R/\lambda$.  $\Delta_1 = 0$
defines a critical
curve separating regions with and without vortices, and  is depicted in
Fig. {\ref{fig:phaseperp}. Stability occurs ($\Delta_N < 0$) for the
regions in Fig. {\ref{fig:phaseperp} below the critical curves. The
asymptotic behavior of $\varepsilon_{mv}$
for large and small values of $R/\lambda$ can be found analytically (see
Appendix C for details):

\begin{eqnarray}
\varepsilon_{mv}  \approx &-& m\phi_0 \;\;\;\;\;\;\;
\bigl(\frac{R}{\lambda }\gg 1\bigr),\\
\varepsilon_{mv}\approx &-& m\phi_0\frac{R}{2\lambda}\;\;\;
\bigl(\frac{R}{\lambda}\ll 1\bigr).
\end{eqnarray}
Thus, asymptotically the curve $\Delta_1=0$ turns into a horizontal
straight line $m\phi_0/\varepsilon_v=1$ at large $R/\lambda$ and
into  a logarithmically distorted hyperbola
$(m\phi_0/\varepsilon_v)(R/\lambda)=2$ at
small $R/\lambda$.

\begin{figure}[htb]
\begin{center}
\includegraphics[angle=270,width=3.5in,totalheight=3.5in]{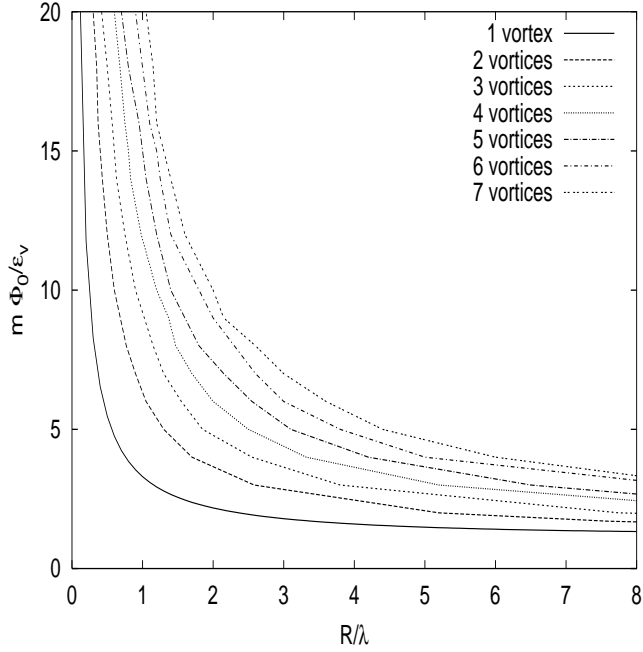}
\caption{\label{fig:phaseperp}Magnetic dot on a superconducting film. }
\end{center}
\end{figure}

For further increase of either $m\phi_0/\varepsilon_v$ or
$R/\lambda$, the second vortex becomes energetically  favorable. Due to
symmetry, the centers of the two vortices are located on a
straight line connecting the vortices with the center of the dot
at equal distances from the center.
The curve 2 on
Fig. \ref{fig:phaseperp} corresponds to this second phase
transition. The occurrence of 2 vortices can be experimentally
detected as the violation of circular symmetry of the field. For
 three
vortices, the equilibrium configuration is a regular
triangle. The further increase of $m\phi_0/\varepsilon_v$ or
$R/\lambda$ makes other vortex states more energetically favorable. 
In principle there exists an infinite
series of
such transitions. Here, we limit ourselselves to the first seven 
transitions by considering the next four vortex states. In equilibrium, 
for vortices sit on the corners of a square, whereas five vortices form a 
pentagon. We find that geometrical pattern for six vortices is hexagon. 
For the case of seven vortices, equilibrium configuration is different 
from the first six states. Namely, one vortex is situated at the dot's 
center, while other six vortices form a hexagonal shape ( see 
Fig.\ref{fig:sub}).

\begin{figure}
\centering
\subfigure[Two vortices] 
{
    \label{fig:sub:a}
    \includegraphics[width=4cm]{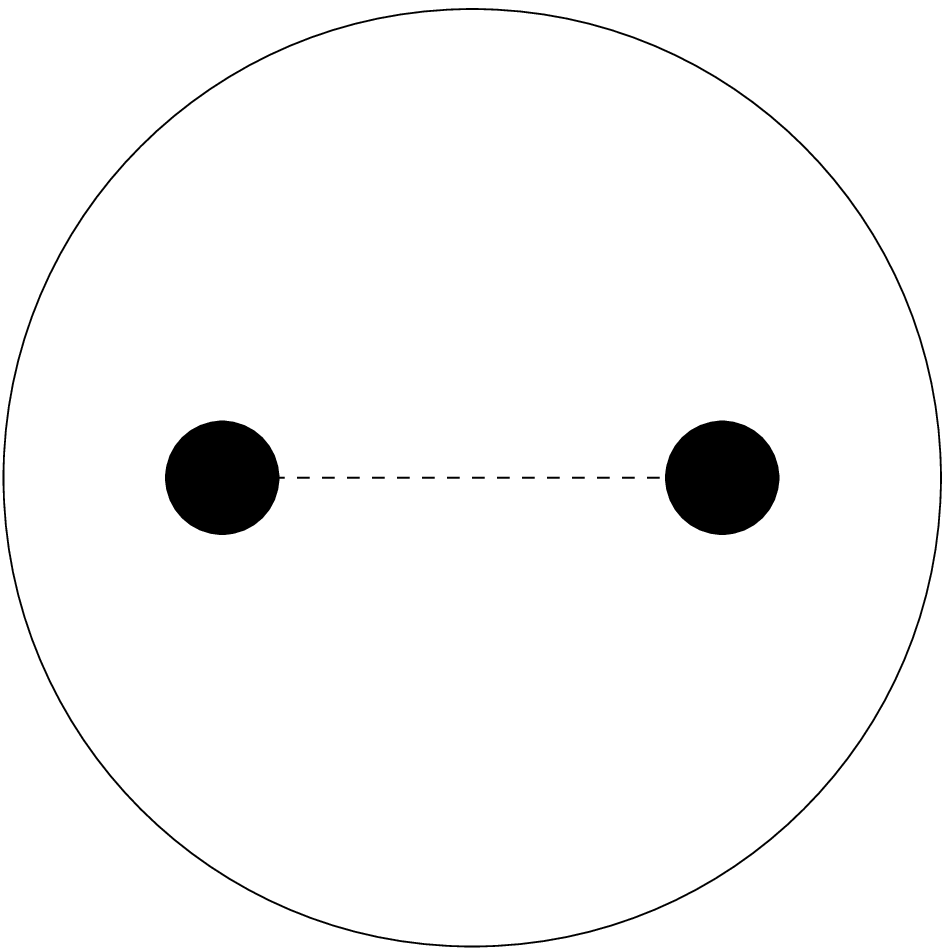}
}
\hspace{1cm}
\subfigure[Three vortices] 
{

    \label{fig:sub:b}
    \includegraphics[width=4cm]{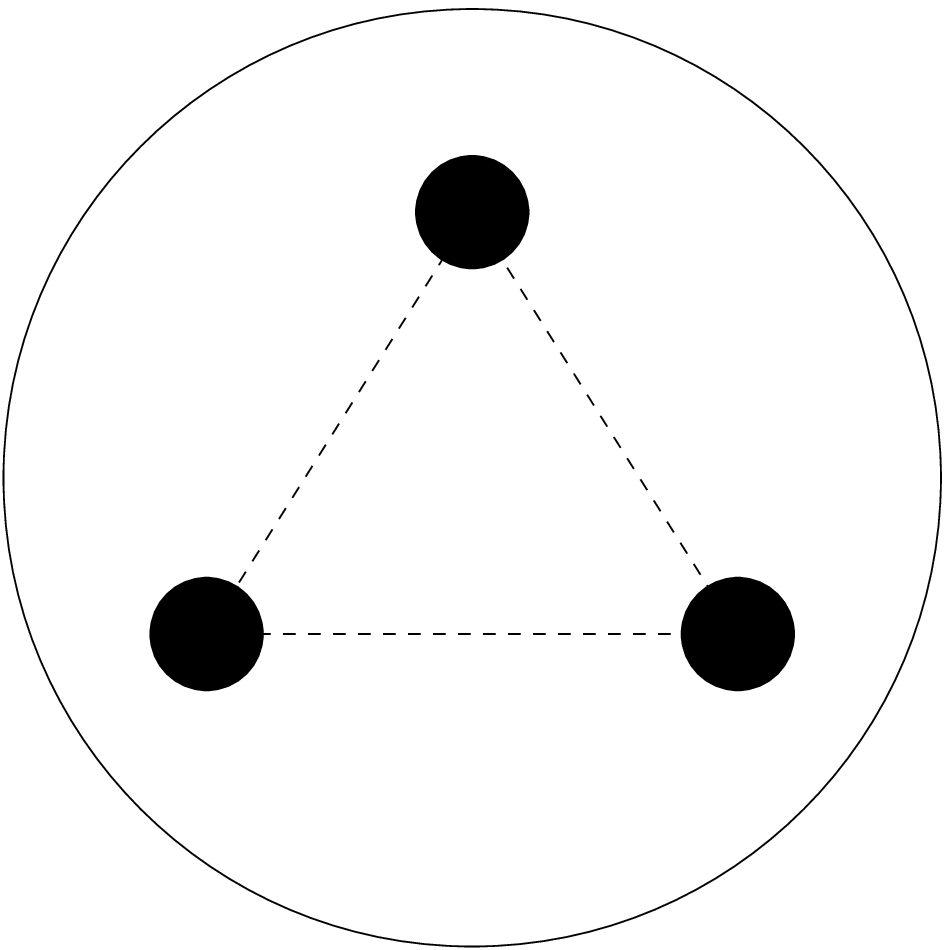}
}
\hspace{1cm}
\subfigure[Four vortices] 
{
    \label{fig:sub:c}
    \includegraphics[width=4cm]{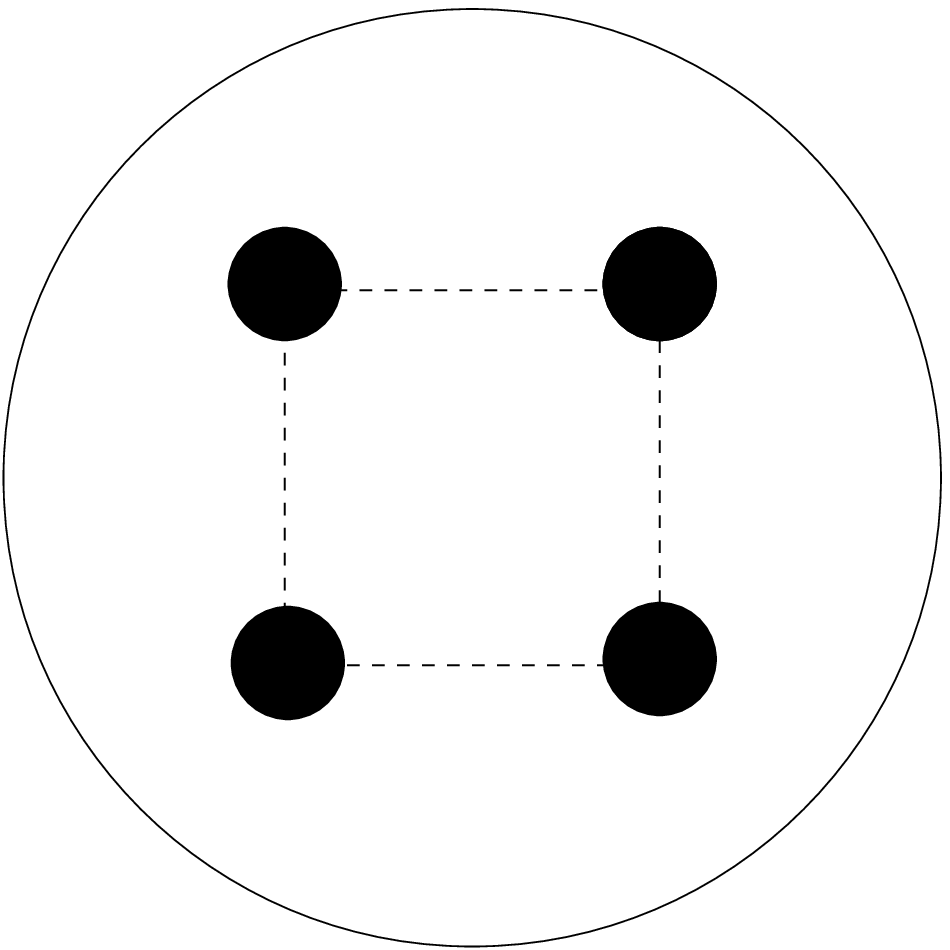}
}
\hspace{1cm}
\subfigure[Five vortices] 
{
    \label{fig:sub:d}
    \includegraphics[width=4cm]{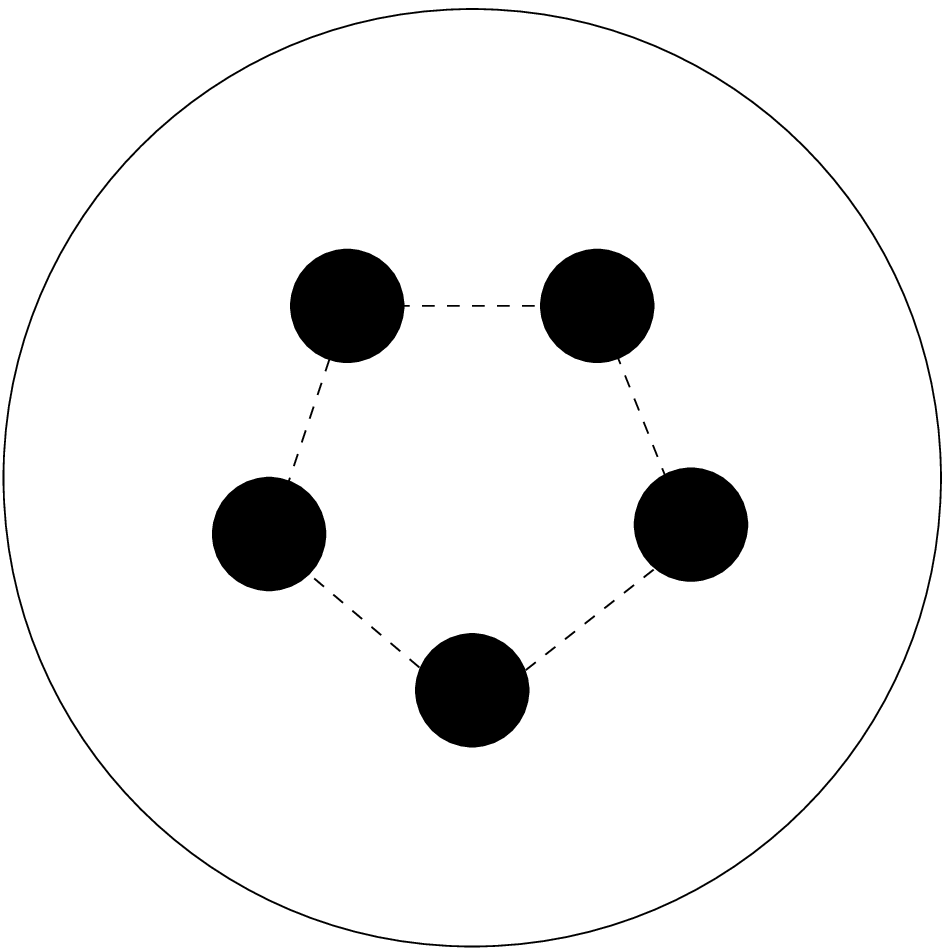}
}
\vspace{1cm}
\subfigure[Six vortices] 
{
    \label{fig:sub:e}
    \includegraphics[width=4cm]{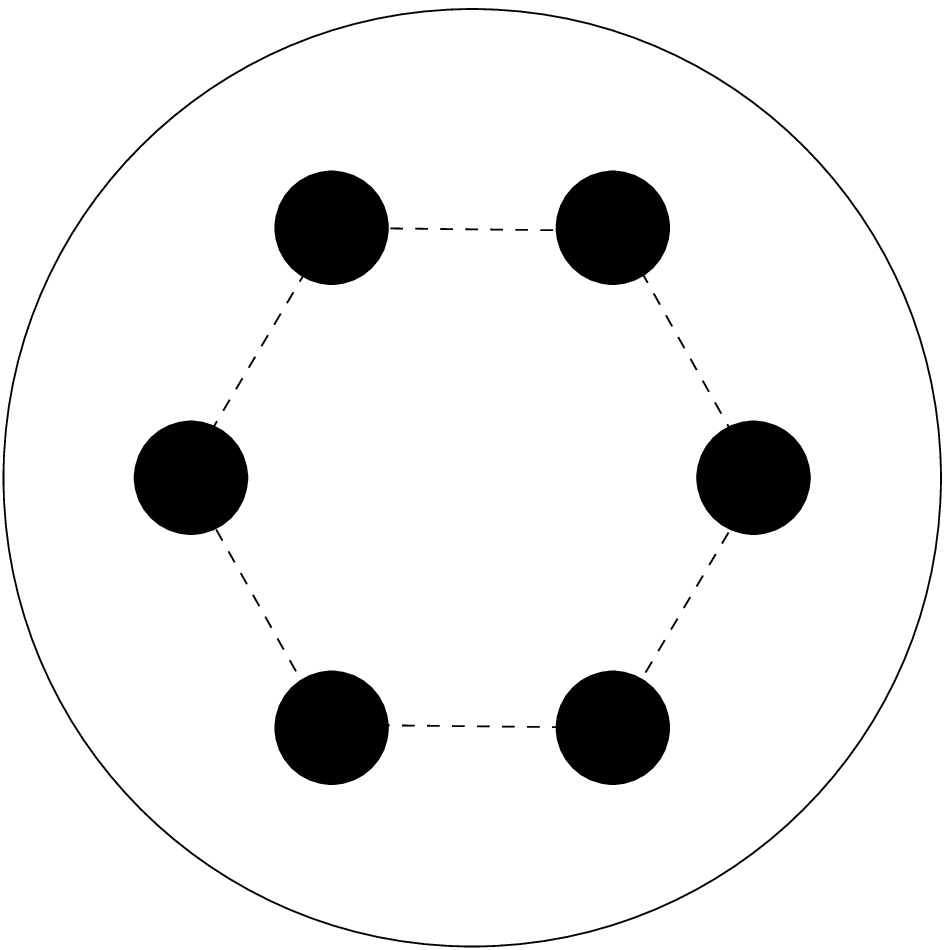}
}
\hspace{0.5cm}
\subfigure[Seven vortices] 
{
    \label{fig:sub:f}
    \includegraphics[width=4cm]{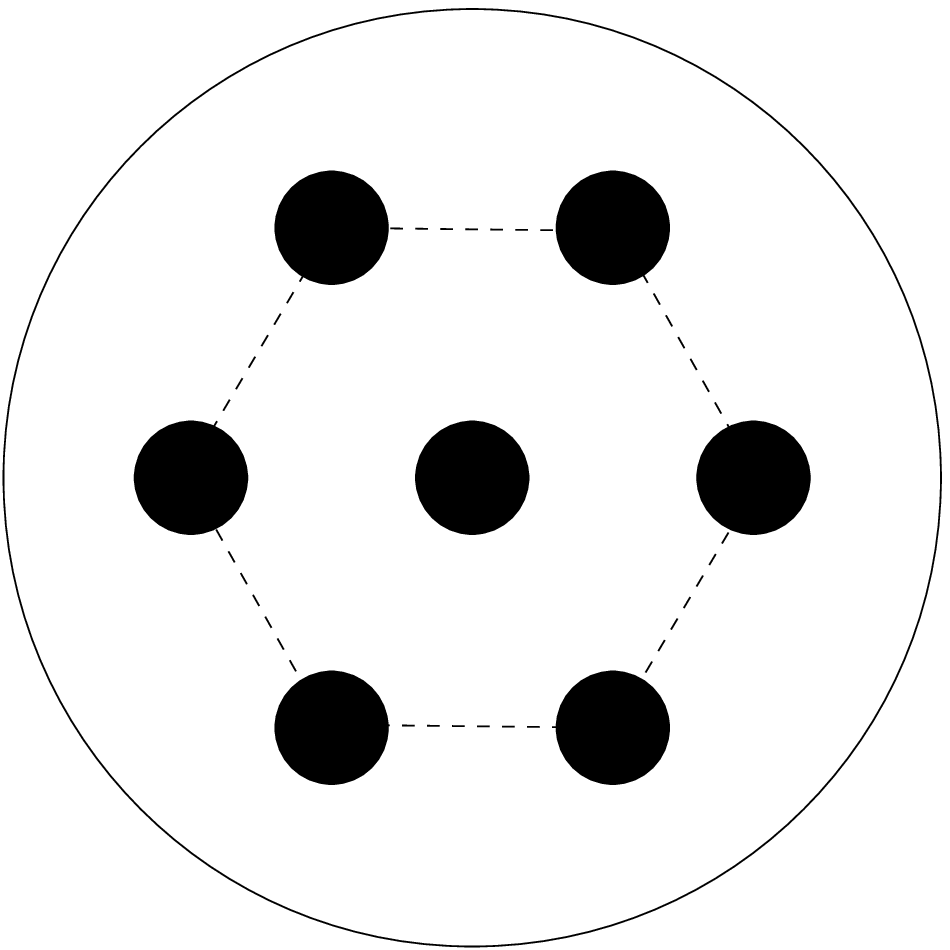}
}
\caption{Vortex states}
\label{fig:sub} 
\end{figure}

It is
not yet
clear what is the
role of configurations with several vortices confined within the
dot region and antivortices outside. Note that these results are valid for 
infinite systems. When the FM dot is placed on top of a finite SC film, we 
suspect that the geometrical patterns formed by vortices will be quite 
different. To investigate this hypothesis, we recently studied the FM dot 
on 
a semi-infinite SC film, and found that the vortex is shifted either 
towards or away from the SC film's boundary due to the competition 
between the attraction of a vortex with the SC film's edge via its image 
vortex and the interaction between the dot and the vortex 
\cite{erdin-semi}.

\section{FERROMAGNETIC-SUPERCONDUCTING BILAYERS}

Earlier
Lyuksyutov and Pokrovsky noticed \cite{pok2,pok7} that in a
bilayer consisting of homogeneous superconducting (SC) and ferromagnetic
(FM) films  with the magnetization
normal to the plane, SC vortices occur spontaneously in the ground state,
even
though
the magnet does not generate a magnetic field in the SC film. In previous 
work \cite{ser3}, we presented  a theory of such
vortex-generation instability and the resulting vortex structures, and  
showed that 
that due to this instability, domains with alternating
magnetization and vortex
directions occur in ferromagnetic-superconducting bilayers (FSB). In 
this chapter, we review the theory of FSB and the domain structures. In the
next section, we treat these domain structures  in the continuum regime
in which the
domain
size is much larger than the  effective penetration depth. This
approximation does not work when the equilibrium size of the domains is
on the order of the effective penetration depth. However, it can be
recovered by considering the
discrete lattice of vortices instead. In the second section, we
report our preliminary results on 
the possible  equilibrium structure in the discrete case  and
calculate the vortex positions, which  depend  on the
magnetization
and
domain   wall energy \cite{domwall}.

\subsection{The Continuum Regime}

We start by refining previous arguments establishing a topological
instability in the FSB \cite{pok2,pok7}. We assume that the magnetic
anisotropy is strong enough to keep the magnetization exactly
perpendicular to the film (in the $z$-direction). The homogeneous
FM film creates no magnetic field outside itself and hence does
not alter the state of the SC film. The
magnetic field generated by a  single
vortex in the superconducting film, with magnetic flux $\phi_0=hc/2e$,
interacts with the
magnetization ${\bf m}$ of the FM film and lowers the total energy
by  $-m\phi_0$ for a proper sign of vorticity. The
energy
to create a Pearl vortex in an isolated SC film is
$\varepsilon_{v} =\varepsilon_0 \ln (\lambda/\xi)$ \cite{pearl},
where $\varepsilon_0 = \phi_0^2/16\pi^2 \lambda$, $\lambda =
\lambda_{L}^{2}/d$ is the effective penetration
depth \cite{abr}, $\lambda_{L}$ is the London penetration
depth, and  $\xi$ is the coherence length. Thus, the total energy
of a single vortex in the FSB is
\begin{equation}
\varepsilon_v^{eff} = \varepsilon_{v} - m \phi_0,
\label{ven}
\end{equation}
\noindent and the FSB becomes unstable with respect to spontaneous
formation vortices as soon as $\varepsilon_v^{eff}$ turns negative.  Note
that close enough to the SC transition temperature $T_s$,
$\varepsilon_v^{eff}$ is definitely negative since the SC electron density
$n_s$ and, therefore, $\varepsilon_{v}$ is zero at $T_s$ (Recall that
$\varepsilon_v (T) = \varepsilon_v (T=0) (1 -\frac{T^2}{T_s^2})$). For
small
$m$ value, $\varepsilon_v^{eff}>0$ at $T=0$, the instability
exists in the temperature interval $T_v<T<T_s$, where
$\varepsilon_v^{eff}(T_v)=0$, otherwise instability persists until $T=0$.

A newly appearing vortex phase cannot consist only of vortices of
one sign. Indeed, any system with average vortex density $n_v$
would generate a constant magnetic field $B_z = n_v \phi_0$ along
the $z$ direction.  The energy of this field for a finite film of
the linear size $L_f$ grows as $L_f^3$, which quickly
exceeds the gain in energy due to creation of vortices,
proportional to $L_f^2$. Hence, in order for the vortex array to
survive, the film should split in domains with alternating
magnetization and vortex directions. We show below that if the
domain size $L$ is much greater than the effective
penetration length $\lambda$, the most favorable arrangement is
the stripe domain structure.
To this end we write the total energy of the bilayer in the form
\begin{equation}
U\,=\,U_{sv}+U_{vv}+U_{mv}+U_{mm}+U_{dw}, \label{energy}
\end{equation}
where $U_{sv}$ is the effective energy of single vortices; $U_{vv}$
is the vortex-vortex interaction energy; $U_{mv}$ is the energy of
interaction between the vortices and magnetic field generated by domain
walls; $U_{mm}$ is the self-interaction energy of the magnetic layer;
and $U_{dw}$
is the linear tension energy of magnetic domain walls \cite{domwall}. We
assume
that the
2d
periodic
domain structure consist of two equivalent sublattices, so that the
magnetization $m_z({\bf r})$ and density of vortices $n({\bf r})$
alternate when crossing from one sublattice to another. The magnetization
is
assumed to have a constant absolute value: $m_z({\bf r})=ms({\bf r})$,
where $s({\bf r})$ is the periodic step function equal to $+$1 at one
sublattice and $-$1 at the other one. We consider a dilute vortex system
where the vortex spacing is much larger than $\lambda$. Then the
effective single-vortex
energy becomes
\begin{equation}
U_{sv}\,=\,\varepsilon_v^{eff}\int n({\bf r})s({\bf r})d^2x.
\label{single}
\end{equation}
\noindent Note that $ n({\bf r})s({\bf r}) > 0$ in all cases. Due to
``average neutrality$^{,,}$ of
the periodic stripe system, the energy of a single vortex in equation
(\ref{single}) is different from
(\ref{ven}) : $\varepsilon_v^{eff} = \varepsilon_v - m \phi_0 / 2$. Note
that,
$-\frac{1}{2}\int {\bf b}^v \cdot {\bf m} d^2 x$ term in Eq.(\ref{en-mv})
contributes $ - m \phi_0 / 2$ in the effective single-vortex energy. In
the periodic systems, the contribution of the surface term is zero. The
vortex-vortex interaction energy is
\begin{equation}
U_{vv}\,=\,\frac{1}{2}\int n({\bf r})V({{\bf r-r}^{\prime}})n({\bf
r}^{\prime})d^2xd^2x^{\prime}, \label{vv}
\end{equation}
where $V({\bf r - r}^{\prime})$ is the pair interaction energy
between vortices located at points ${\bf r}$ and ${\bf r}^{\prime}$.  Its 
asymptotic value at large distances $\mid{\bf r -r}^{\prime}\mid\gg \lambda$  
is \cite{degennes}

\begin{equation}
 V({\bf r - r}^{\prime})=\frac{\phi_0^2}{4\pi^2\mid{\bf r -
r}^{\prime}\mid}.
\end{equation}
This long-range interaction is induced by
the magnetic field generated by the Pearl vortices and their slowly
decaying currents. By Eq.(\ref{en-mv}), the energy of the vortex
interaction with the
magnetic field generated by the magnetic film is
\cite{erdin-common}
\begin{equation}
U_{mv}\,=\,-\frac{\phi_0}{16 \pi^2 \lambda}\int \nabla \varphi ({\bf r} -
{\bf
r^\prime}) n({\bf r}^\prime)\cdot {\bf a}^{(m)}({\bf 
r})d^2xd^2x^{\prime}.\label{vm}
\end{equation}
Here $\varphi ({\bf r} - {\bf
r^\prime})=\arctan\frac{y-y^{\prime}}{x-x^{\prime}}$ is a phase
shift created at a point ${\bf r}$ by a vortex centered at a point
${\bf r}^{\prime}$ and ${\bf a}^{(m)}({\bf r})$ is the value of
the vector-potential induced by the FM film upon the SC one. By
Eq.(\ref{en-m}), the
magnetic self-interaction reads
\begin{equation}
U_{mm}\,=\,-\frac{m}{2}\int B_z^{(m)}({\bf r})s({\bf r}) d^2x.
\label{mm}
\end{equation}
Finally, each  magnetic domain wall's linear energy is
$U_{dw}=\varepsilon_{dw}L_{dw}$, where $\varepsilon_{dw}$ is the linear
tension of the magnetic domain wall and $L_{dw}$ is the total length of
the
magnetic domain walls.
Let us analyze the vortex-domain-wall interaction $U_{mv}$.
The magnetic vector-potential ${\bf A}^{(m)}$ obeys the London-Pearl
magneto-static equation (see Eq.(\ref{vec})):

\begin{equation}
\nabla \times\left(\nabla\times{\bf A}^{(m)}\right)=
\left[ -\frac{1}{\lambda }{\bf a}^{(m)}{\bf +}
4\pi \nabla \times (\widehat{z}m({\bf r}))\right]
\delta (z).
\label{LP}
\end{equation}
We consider $L\gg \lambda$, where the term
$\nabla \times\left(\nabla\times{\bf A}^{(m)}\right)$
is negligible and then
\begin{equation}
{\bf a}^{(m)}\approx -4\pi m\lambda \hat{z}\times \nabla s({\bf r}).
\label{a-m}
\end{equation}

The phase gradient entering Eq. (\ref{vm}) can be rewritten as:
$\nabla \varphi ({\bf r})=\hat{z}\times \nabla \ln |{\bf
r-r}^{\prime }|$. Plugging this expression into (\ref{vm}),
integrating by part and employing relation\\
 $\nabla ^{2}\ln |{\bf
r-r}^{\prime }|=-2\pi \delta ({\bf r-r}^{\prime })$, we arrive at
\begin{equation}
U_{mv}\,=\,-\frac{\phi_{0}}{2}\int m({\bf r})n({\bf r})d^{2}x.
\label{vm-res}
\end{equation}
This result implies that the vortex-domain-wall interaction
renormalizes the single-vortex to
\begin{equation}
\tilde{\varepsilon}_v=\varepsilon_v- m\phi_0.
\end{equation}
Thus, the term
$U_{mv}$
can be removed from the total energy (\ref{energy}) if the
single-vortex contribution $U_{sv}$ is replaced by
$\tilde{U}_v$, which differs drom  (\ref{single}) on  replacing
$\varepsilon_v^{eff}$ by $\tilde{\varepsilon}_v$. In physical terms, it
means
that the vortex attraction to the magnetic domain walls lowers the
threshold for the spontaneous appearance of the vortex-domain
structure. The next step is the minimization of energy with respect to 
$n({\bf r})$
the vortex
density, which appears only in the first three terms of the total
energy (see Eq.(\ref{energy})). Their sum can be conveniently denoted by
$U_v\equiv\tilde{U}_{sv}+U_{vv}$. To simplify the minimization, we
 Fourier-expand  the periodic functions: $s({\bf
r})=\sum_{\bf G} s_{\bf G}e^{i{\bf G}\cdot{\bf r}}$ and $n({\bf
r})=\sum_{\bf
G} n_{\bf G}e^{i{\bf G}\cdot{\bf r}}$. The energy $U_v$ in
the Fourier-representation then reads
\begin{equation}
U_v = U_{sv} + U_{vv} + U_{mv} 
= \sum_{\bf G}\left(\tilde{\varepsilon}_vs_{\bf G}n_{-{\bf G}}+
\frac{1}{2}V_{\bf G}n_{\bf G}n_{-{\bf G}}\right),
\label{U-v-F}
\end{equation}
where $V_{\bf G}=\int V({\bf r})e^{i{\bf G}\cdot{\bf r}}d^2x=
\phi_0^2/2\pi
|{\bf G}|$. Minimization of Eq.(\ref{U-v-F}) over $n_{\bf G}$ leads
to
\begin{eqnarray}
n_{\bf G}\,=\,-\frac{\tilde{\varepsilon}_v s_{\bf G}}{V_{\bf G}}\,=\,
-\frac{2\pi\tilde{\varepsilon}_v|{\bf G}|s_{\bf G}}{\phi_0^2},
\label{dens-F}\\
U_v\,=\,-\frac{\pi\tilde{\varepsilon}_v^2}{\phi_0^2}\sum_{\bf G}|{\bf
G}||s_{\bf G}|^2. \label{U-v-eq}
\end{eqnarray}
Note that the solution becomes physically meaningless at positive
$\tilde{\epsilon}_v$.
We now apply these general results to analyze the stripe domain
structure. In this case the density of vortices $n(x)$ depends
only on one coordinate $x$ perpendicular to the magnetic domain walls. The
vectors ${\bf G}$ are directed along the $x$-axis. The allowed
wave numbers are $G=\pi (2r+1)/L$, where $L$ is the domain
width and $r$ runs over all integers. The Fourier-transform of the step
function is
$s_G=\frac{2i}{\pi (2r+1)}$. The inverse Fourier-transform of Eq.
(\ref{dens-F}) for the stripe domain case is (see Appendix C for details)
\begin{equation}
n(x)\,=\, -\frac{4 \pi \tilde{\varepsilon}_v}{\phi_0^2 L}
\frac{1}{\sin\frac{\pi x}{L}}.
\label{dens-x}
\end{equation}
Note the strong singularity of the density near the domain
walls. Our
approximation is invalid at distances of the order of $\lambda$,
and the singularities must be smeared out in a band of the width
$\lambda$ around the magnetic domain wall. Conversely, the approximation
of
the zero-width magnetic domain wall is invalid within the range of
the magnetic domain wall width $l$. Fortunately, we do not need more
detailed information on the distribution of vortices in the vicinity of 
the
magnetic domain walls. Indeed, by substituting the Fourier-transform of
the
step function into equation (\ref{U-v-eq}), we find
a logarithmically divergent series in the form of $\sum_r 1/(2r+1)$. It
must be cut off at $\pm
r_{max}$ with $r_{max}\sim L/\lambda$. The summation can be
performed using the Euler asymptotic formula \cite{GR} with the
following result (see Appendix C for details):
\begin{equation}
U_v^{str}=-\frac{4\tilde{m}^{2}{\cal A}}{L}\left( \ln
\frac{L}{\lambda }+C+2\ln2\right), \label{stripe-u}
\end{equation}
where $\tilde{m}=m-\varepsilon_v/\phi_0$, ${\cal A}$ is the domain area
and $C\sim
0.577$. Now the
problem is to analyze the
proper cut-off for any lattice. As we have seen
in the stripe domain structure, the energy $U_v$ diverges logarithmically
due to the strong
singularity of the vortex density near each magnetic domain wall (see Eq.
(\ref{dens-x})). Thus, the logarithmic term is proportional to the
total magnetic domain wall length (see Eq.(\ref{stripe-u})). The
singularity of the vortex density contributes the similar logarithmic term
to the energy for any lattice. However, this logarithmic accuracy is
not sufficient to distinguish the domain structures. In order to
solve this problem, we need the
next
approximation to the energy $U_v$, i.e. a
term $\alpha$, proportional to the length of the magnetic domain wall,
without the
logarithmic
factor. Together with this term, the energy for any lattice looks like
$U_v \sim \ln ( L \alpha/\lambda)$. Now, the problem is to find this
term accurately. Such a term includes a non-local contribution from
large distances between $\lambda$ and $L$ and a local contribution from
the vicinity of the magnetic domain walls. The non-local contribution is
accurately accounted for by the summation over the integers;
whereas, the local contribution requires a cut-off at large $r$,
which is not well defined. However, due to its local character it
must be the same for all magnetic domain walls. Therefore, it is possible
to choose the maximal wave-vector in the direction normal to the
magnetic domain wall as  $2\pi /\lambda $. Such a procedure
renormalizes the magnetic domain wall's linear tension, in
the same for any domain lattice. This remark allows calculation of
the energy $U_v$ for the square and triangular lattices.
For the square checkerboard lattice, the allowed
wave-vectors are ${\bf G}
=\frac{\pi}{L}\left[(2r+1)\hat{x}+(2s+1)\hat{y}\right]$. The
Fourier-transform of the step function is: $s_{\bf
G}=\frac{4}{\pi^{2}(2r+1)(2s+1)}$. The maximal values of $r$ and
$s$ are identical and equal to $L/\lambda$ where $L$ is the side
of a square domain. The summation, similar to the case of stripe
structure although somewhat more complicated, leads to the
following expression (see Appendix C for details):
\begin{equation}
U_{v}^{sq} = -\frac{8\tilde{m}^{2}{\cal A}}{L}\left( \ln
\frac{L}{\lambda }+C+2\ln 2 -\gamma\right),
\label{U-sq}
\end{equation}
where the numerical constant $\gamma$ is defined below:
\begin{equation}
\gamma=(2-\sqrt{2})\frac{7}{\pi ^{2}}\zeta (3)
+\frac{16}{\pi ^{2}}\sum_{r=0}^{\infty }\sum_{s=r+1}^{\infty }
{\cal S}(r,s).
\label{gamma}
\end{equation}
Here $\zeta (x)$ is the Riemann zeta-function; $\zeta (3)\approx
1.2020$ and,
\begin{equation}
 {\cal S}(r,s)=\frac{ 2(r+s+1)-\sqrt{(2r+1)^{2}+(2s+1)^{2}}}{
(2r+1)^{2}(2s+1)^{2}}.
\end{equation}
Direct numerical calculation gives $\gamma\approx 0.9>\ln
2$.

The reciprocal lattice vectors for the regular triangular domain
lattice are \\
${\bf G=}\frac{2\pi }{L}
\left[ r\left( \widehat{x}-\frac{1}{\sqrt{3}}\widehat{y}\right)
+s\frac{2}{\sqrt{3}}\widehat{y}\right]$. The analysis is
remarkably simplified in the ``triangular coordinate frame":
$u=x-y/\sqrt{3};\,\,v=2y/\sqrt{3}$. The step function inside one
elementary cell is $s(u,v)=+1$ for $u+v<L$ and $s(u,v)=-1$ for
$u+v>L$, where $L$ is the side of the elementary triangle. The
Fourier-transform of the step function $s_{\bf G}$ is not zero at
either $r\neq 0, s=0$, or $r=0, s\neq 0$, or $r=s\neq 0$. For all
these cases $|s_{{\bf G}}|^{2}= 1/(\pi^2 q^2)$, where $q$ is
either $r$ or $s$, depending on which of these numbers differs
from zero. For this case, the summation in equation (\ref{U-v-eq}) gives
\begin{equation}
U_{v}^{tri}=-\frac{32{\cal A}\widetilde{m}^{2}}{L\sqrt{3}}
\left( \ln r_{\max }+C\right).   \label{U-ln-tr}
\end{equation}
However, the value $r_{\max }$ is different from the stripe and
square cases since the coordinates are skewed, here it equals
$\frac{\sqrt{3}}{2}\frac{L}{\lambda }$.

Our next step is to show that the magnetization self-interaction
can be included into the renormalized magnetic domain wall linear tension.
For the isolated FM stripe domain structure, the magnetization
self-interaction energy is
equal to $U_{mm}=-m^{2}L_{dw}\ln \frac{L}{l}$, where $l$ is the
magnetic domain wall width \cite{abanov1,abanov2,abanov3}. The
superconducting
screening
enhances the magnetic field very near the magnetic domain walls and
reduces it
way from the  magnetic domain walls. In the stripe geometry, from
Eq.(\ref{a-m}),
$b_{z}^{(m)}= da^{(m)}/dx =-4\pi m\lambda (d^{2}s/dx^{2})$
implying that the screened magnetic field is confined to an
interval $\sim \lambda $ near magnetic domain walls. Thus, its
contribution
to the energy does not contain a large logarithm. By
(\ref{mm}), $U_{mm}$ gives
$-L_{dw}m^{2}$, which can be incorporated into the renormalized value
of the magnetic domain wall linear tension. Note that this contribution is
negative. We assume that it is less than the initial positive
linear tension $\varepsilon_{dw}$. We do not consider here the
interesting but less likely possibility of a negative
renormalized linear tension, which probably results in  domain
wall branching.

Now we are in position to minimize the total energy $U$ over the
domain width $L$ and compare the equilibrium energy.
The equilibrium domain width and the equilibrium energy
for the stripe structure are
\begin{equation}
L_{eq}^{(str)}=\frac{\lambda }{4}\exp
\left( \frac{\varepsilon _{dw}}
{4\tilde{m}^{2}}-C+1\right),
\label{L-eq-stripe}
\end{equation}
\begin{equation}
{U_{eq}^{(str)}}
=-\frac{16\widetilde{m}^{2}
{\cal A}}{\lambda }\exp
\left( -\frac{\varepsilon _{dw}}{4\widetilde{m}^{2}}
+C-1\right).
\label{U-eq-stripe}
\end{equation}
Calculating the corresponding  values for the square and triangular
lattice,
we obtain:
$L_{eq}^{(sq)}=
L_{eq}^{(str)}\exp\left(\gamma \right)$;
$U_{eq}^{(sq)}=2{U_{eq}^{(str)}}\exp\left(-\gamma \right)$;
and $U_{eq}^{(tri)}=(3/4){U_{eq}^{(str)}}$.
Comparing these energies to the energy
of the stripe structure, for which $U_{eq}^{(str)} < 0$, we conclude that
the
stripe structure wins.
The domains become infinitely wide at $T=T_s$ and at $T=T_v$. The
expression in the exponent (\ref{U-eq-stripe})  at $T=T_s$ is four
times less than the corresponding expression for domains in an
isolated magnetic film \cite{landau}. Therefore, stripe domains in the
bilayer can
be energetically favorable even if the isolated magnetic film
remains in a monodomain state. If stripe domains in the magnetic film
exist above the SC transition, then they shrink dramatically below the
transition. The physical reason behind this effect is as follows. There 
are two contributions to the energy that determines the domain width; 
magnetic energy and the domain wall energy. While the latter contribution 
prefers the larger domain width, the former contribution causes the 
smaller width. In the presence of superconductor, the magnetic  field is 
screened due to the Meissner effect, which makes the magnetic energy 
contribution dominant. As a result, the domain width is smaller than that 
in the case of ferromagnet only. Bulaevsky and Chudnovsky \cite{bul2} 
found that the
domain width in a {\it thick} magnetic layer above  a {\it
bulk} superconductor is proportional to $d_m^{1/3}$ instead of
$d_m^{1/2}$, a well-known result for an isolated magnetic layer.
Here $d_m$ is the thickness of the magnetic layer; "thick"
means that $d_m >> l$.
Our problem is fundamentally different on two counts: first, we
consider a {\it thin} FM film $d_m\ll l$ above a {\it thin} SC
film and, second, the main effect is due to  the interaction of
vortices with the magnetization rather than from the screening of
the magnetic
field  as in  \cite{bul2}. The vortex-magnetization interaction  effect is
much
stronger, leading to a totally different dependence.

\subsection{The Discrete Regime}

If $\varepsilon_{dw}\leq 4\tilde{m}^2$, the continuum approximation
becomes invalid, since $L_{eq}$ becomes on the order of or less
than $\lambda$ (see Eq.(\ref{L-eq-stripe})). Instead a lattice of discrete
vortices must be
considered. In this section, we present a method which works in both
continuum and the discrete regimes. We study the lattices of discrete
vortices only in the stripe phase, since the system favors that phase in
equilibrium.  In the continuum approximation,
it
is found that the vortex density increases at the closer distances to the
magnetic domain walls. Based on this fact and the symmetry of the stripe
domain
structure, it is reasonable to consider that the vortices and
antivortices form periodic structures on
straight  chains along the $y$ direction. Even though it is not clear
how many chains are associated with each domain, we can still make
progress
toward  understanding  discrete vortex lattices. In this section, we 
report our prelimanary results on 
the problems; i) how the vortices and the antivortices are positioned  on 
the
chains; ii) how the equilibrium domain size changes, depending on the
magnetization and the magnetic domain wall energy in the presence of the
vortices. In order to
solve these problems, we propose a configuration of the
vortex
and the antivortex chains in which  two chains per stripe is
considered.  According to 
our 
results in the continuum regime,  
the vortex 
density increases near the domain walls. Therefore, we can have at least 
two chains per stripe in a possible configuration.
Chains are situated at distance $a$ 
from the 
magnetic domain 
walls. Another problem is how to place vortices and antivortices on the 
chains. If vortices are  next to
each other on the either side of the magnetic domain wall, the magnetic
fields
they produce   cancel out each other. As a result, the gain in
energy is   diminished. However, the system has the largest gain when
the  vortices and antivortices are shifted by a half period 
$b/2$ (see Fig. \ref{2col}).

\begin{figure}[t]
\begin{center}
\includegraphics[angle=0,width=3.5in,totalheight=3.5in]{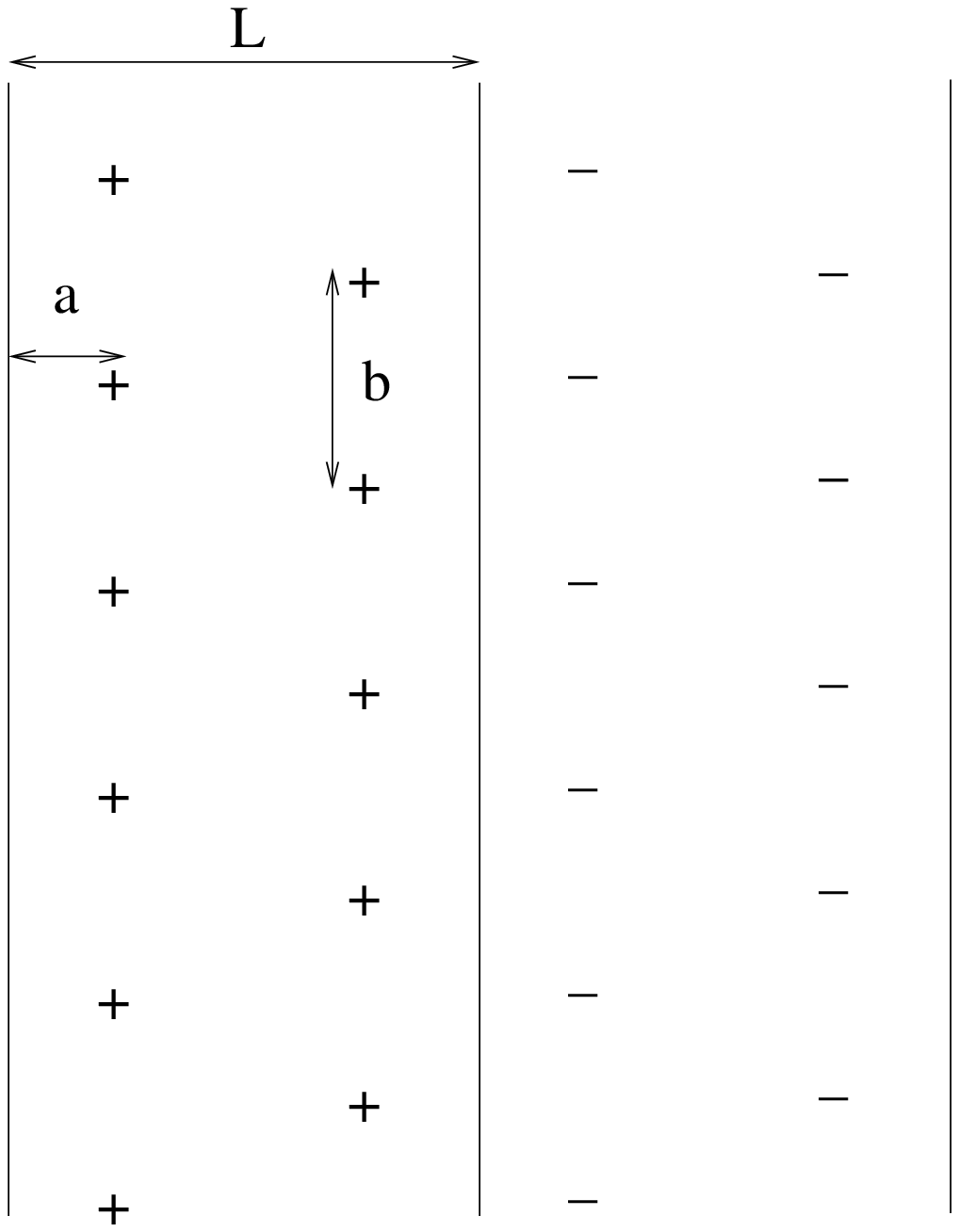}
\caption{The vortex lattice. \label{2col}}
\end{center}
\end{figure}

Our next step is to write the energies of  the proposed structure. To this
end, we use (\ref{hv},\ref{hmv},\ref{hmm}). In those equations, the vortex
configurations  differ
by their form-factors. We can obtain them from $F_{\bf G} =
\sum_{{\bf r}_i} n_i e^{i {\bf G} \cdot {\bf r}_i}$, where the ${\bf G}$'s
are
the
reciprocal vectors of the periodic structures, the ${\bf r}_i$ are the
positions
of the vortex centers,  and $n_i$ are  the charge of the vortex. In our
proposed
model,
${\bf G} =  ( ( 2 r + 1 )
\frac{\pi}{L}, 2 s \frac{\pi}{b})$, $n_i = \pm 1$ and $F_{\bf G} = e^{i 
G_x a} - (-1)^ s e^{- i G_x a}$. In our calculations, the
divergent part of the series must be
extracted carefully. We show below the detailed  analysis of series
equations for each candidate. We start with the
self interaction energy of the magnetic layer $U_{mm}$, since it is the
same for each configuration. For the periodic structures, it is given by
({\ref{hmm}). Direct substitution of the Fourier coefficient of the
stripe phase $m_{z{\bf G}}=\frac{2 i
m }{\pi( 2 r + 1)}$ into Eq.({\ref{hmm}) gives the self-interaction of
the magnetic layer per unit cell as 
\begin{equation}
u_{mm} = - \frac{8 m^2}{L} \sum_{r=0}^{\infty} \frac{1}{\frac{L}{2 \pi
\lambda} + 2 r + 1}.
\label{Hmmser}
\end{equation}

\noindent This series is logarithmically divergent. However, it can be
pulled out easily by adding and substracting $1/(2 r +1)$ in the series
above. Thus, we get two terms; one convergent, the other divergent.
Summing
over $r$ on the divergent part up to the cutoff $r_{max} = L/l$, where
$l$ is the magnetic domain wall width, we obtain the following:

\begin{equation}
u_{mm} = -  \frac{4 m^2}{L}\left(\ln \frac{L}{l} - \psi^{(0)}
(\frac{1}{2}+\frac{L}{ 4 \pi
\lambda})\right),
\label{hmmstr}
\end{equation}

\noindent where $\psi^{(0)} (x)$ is the polygamma function of zeroth
order \cite{abrom}.  In our numerical calculations, we  write the
logarithmic term in (\ref{hmmstr}) as $\ln (\lambda/l) + \ln (L/\lambda)$
and then incorporate the  $-4m^2 \ln (\lambda/l)$ term in
the renormalized $\varepsilon_{dw}^{ren}$.
Another energy term with a divergent series is the
vortex energy, in
general given by (\ref{hv}). The logarithmic divergence in this term stems
from the vortex self-energies.
We first split (\ref{hv})  into two
parts  as follows:
\begin{equation}
u_{vv} = \frac{\pi \varepsilon_0}{2 L^2 b^2} \sum_{\bf G}\left[ \frac{
|F_{\bf
G}|^2}{G^2} -
\frac{ |F_{\bf G}|^2}{G^2 (1 + 2 \lambda G)}\right].
\label{vorener}
\end{equation}

\noindent Note that the area of the elementary cell  is $2 L b$. The first
term of the series above contributes to the self-energies of the
vortices; whereas, the second term is the vortex-vortex energy
and will be left in the series form.
The series in the first term can be transformed to the
form of 
$\sum_{r=-\infty}^{\infty}\sum_{s=-\infty}^{\infty}
1/((2 r + 1)^2 x^2 + s^2)$ where $x$ is constant, and depends on the
form-factor. A detailed analysis of such series is given  in Appendix
B.

The next step is to find the vortex
energy  and the interaction energy of the magnetization and vortices
 for each configuration. In the calculation of $u_{mv}$, we take the
Fourier coefficient of the magnetization to be  $\frac{4 i
m }{( 2 r + 1)}\delta (G_y)$. The fact that the stripe is infinite along
the $y$
direction results in the additional term $2 \pi \delta (G_y)$. However, it
does
not play any role in the calculation of $u_{mm}$. For
numerical analysis, these energies
must be expressed in terms of dimensionless parameters. To this end,  we
define dimensionless variables
$\tilde \lambda = \lambda/L$ , $\tilde b = b/L$ and $\tilde
\varepsilon_{dw}
 =  \varepsilon_{dw}^{ren} \lambda/\varepsilon_0$.  The total energy
$\tilde U$
is
measured in units of $\varepsilon_0/\lambda^2$. In addition, we
introduce the dimensionless magnetic energy as $\tilde U_{mm} =
u_{mm}/(\varepsilon_0/\lambda^2)$. 
 In the fourth configuration, the square of the
form-factor is: \\
$|F_{\bf G}|^2 = 2 - 2 (-1)^s \cos ( ( 2 r + 1 ) \pi
\tilde a )$. Even and odd values of $s$ give different contributions.
Then, we can  calculate the vortex energy for even $s$ and odd $s$
separately. In terms of these parameters, we find

\begin{equation}
\tilde U = \frac{\tilde \lambda^2}{2 \tilde b} \left( \ln (
\frac{\lambda}{\tilde \lambda \xi}) -   2 f_v ( \tilde
\lambda, \tilde a)
-
\frac{4}{\tilde b \pi} f_{vv} ( \tilde \lambda, \tilde a, \tilde b
) -
\frac{16m
\phi_0}{\varepsilon_0} f_{mv} (\tilde \lambda )\right) + \tilde
U_{mm} +
\tilde
\epsilon_{dw} \tilde \lambda,
\label{conf4}
\end{equation}

\noindent where 
\begin{eqnarray}
f_v &=& \sum_{r=0}^{\infty} \frac{\coth (( 2 r + 1
)\frac{\pi \tilde b}{4} ) -1}{2 r + 1} \sin^2 (( 2 r + 1 ) \pi \tilde a)
\nonumber
\\
 &+& \sum_{r=0}^{\infty} \frac{\tanh (( 2 r + 1
)\frac{\pi \tilde b}{4} ) -1}{2 r + 1} \sin^2 (( 2 r + 1 ) \pi \tilde a),
\nonumber \\
f_{vv} &=&  \sum_{r,s = - \infty}^{\infty} \frac{\sin^2 (( 2 r + 1 )
\pi
\tilde a)}{(( 2
r +1 )^2 + \frac{16 s^2}{\tilde b^2})(1+ 2 \pi \tilde \lambda
\sqrt{( 2
r +1 )^2 + \frac{16 s^2 }{\tilde b^2}})}, \nonumber \\
 &+&  \sum_{r,s = - \infty}^{\infty}\frac{\cos^2
(( 2 r + 1 ) \pi
\tilde a)}{ (( 2
r +1 )^2 + \frac{4( 2 s+ 1)^2}{\tilde b^2}) (1+ 2 \pi \tilde \lambda
\sqrt{( 2
r +1 )^2 + \frac{4 (2 s +1)^2    }{\tilde b^2}})}, \nonumber \\
f_{mv} &=& \sum_{r=0}^{\infty} \frac{\sin (( 2 r + 1 )
\pi \tilde a)}{(2 r + 1 ) (1+2 \pi \tilde \lambda ( 2 r + 1 ))}.
\end{eqnarray}

In the numerical minimization of
Eq.(\ref{conf4}), we
take
$\ln(\lambda/\xi) = 5$. Changing  $m
\phi_0/\varepsilon_0$ at fixed $\tilde
\varepsilon_{dw}$, which initially  is fixed at $0.5$,  we
calculate the
minimal energy of the proposed configuration.
We first investigate when this configurations becomes energetically
favorable in the system. To this end, we check where the equilibrium
energies of the configuration first become negative.

\begin{table}[h]
\caption{The numerical results for the fourth configuration at  $\tilde
\varepsilon_{dw}=0.5$. The column on the left is input. \label{disc}}
\begin{center}
\begin{tabular}{|c|c|c|c|}\hline
{$m\phi_0/\varepsilon_v$}&{$L/\lambda$}&{$a/\lambda$}&{$b/\lambda$}\\ 
\hline
 2.00 & 2.50 & 0.56 &0.18  \\ \hline
 3.00 & 1.67 & 0.38 & 0.12  \\ \hline
\end{tabular}
\end{center}
\end{table}

In numerical calculations, we also found that the vortex lattice is stable
for
$\varepsilon_{dw} >4 \tilde
m^2$. At this point, the domain size is noticably larger than the
effective penetration depth $\lambda$, so  the continuum approximation is
valid. Therefore, we
expect that the domain nucleation starts in the continuum
regime. This problem is left for the future research. As seen in
Table \ref{disc}, at constant $\tilde \varepsilon_{dw}$, with
increasing $m \phi_0 / \varepsilon_v$, the equilibrium size of the domain
decreases. In addition, the vortices on the chain get closer to each
other. These results agree with those obtained in the continuum 
approximation.  As
$\varepsilon_{dw}/4\tilde{m}^2$ increases, we expect that new vortex
chains
develop within the domains.
We leave this problem
to future research.

\section{Conclusions}

We reviewed theory of   the heterogeneous
magnetic superconducting systems (HMSS) based on London-Maxwell equations 
and the application of the theory on two realizations: 
ferromagnetic (FM) dots and
their square
array on a superconducting (SC) film, and
ferromagnetic-superconducting bilayers (FSB).
In the first chapter,  we presented a general
formalism for the
interaction between magnetic textures and superconductors in the
Londons approximation. The problem is formulated as a variational
principle. The variational functional (energy) is an integral over
regions occupied either by magnet or by superconductor. It allows
us to find directly the positions of vortices and magnetization.
Afterwards, the formalism is extended to the case of periodic 
structures and finite systems.

As applications of the formalism, we have shown that vortices in
superconducting
films can be generated by magnetic dots magnetized normal to the
film. We
have found phase transition curves separating
the state without vortices from the state with one vortex and the
latter from the state with two vortices.
Up to 7 vortices, the vortex-configurations on the ground state of a 
SC
film with the FM dot on it are determined.
For one vortex
under a dot we have shown that the perpendicular component of
magnetic field changes sign at some distance from the dot. This
fact can be used for diagnostics of the vortex generation.

However, we treat only vortices under the FM dot. In a more realistic
picture, the antivortices outside the dot  become important  and
most likely affect the configurations of the vortices confined within
the dot's region. This problem still remains open.

In the fourth chapter, we studied ferromagnetic-superconducting 
bilayers
(FSB). We predicted that in a finite
temperature interval
below the SC
transition the
FSB is unstable with respect to SC vortex formation.  The slow decay
($\propto 1/r$) of the long-range interactions between Pearl
vortices makes the structure that consists of alternating domains 
with
opposite magnetization and vorticity energetically favorable. The
distribution of vortices inside each domain is highly inhomogeneous, 
with
density increasing near the magnetic domain walls. As
long as the
domain width
is larger than the effective penetration depth, the energy of the 
stripe
domain structure is minimized. These new topological structures can 
be
observed directly. A strong anisotropy in current  transport would 
provide
indirect evidence of the stripe texture: the bilayer may be
superconducting for current parallel to the domains and resistive for
 current perpendicular to the domains.

If $\varepsilon_{dw}\leq 4\tilde{m}^2$, the continuum approximation
becomes invalid. Instead, we considered the discrete lattice of 
vortices. We analyzed
the vortex configurations in  two vortex chains.
Depending on the magnetization and the
magnetic domain wall energy, the positions of the vortices and the
equilbrium
domain
size are calculated.

It is possible that the long domain nucleation time can interfere 
with
the observation of described textures. We expect, however,  that the
vortices that appear first will reduce the barriers for domain
walls and, subsequently, expedite domain nucleation.
Quantitative study of this dynamic process is still in progress.

Our purpose was to consider quantitatively a new class of phenomena
provided by the interaction between superconductivity and 
ferromagnetism
in
heterogeneous systems. This review article  focuses  on unusual 
equilibrium
structures of vortices and magnetization occuring in such systems. As
the simplest specific examples we considered single magnetic dots and 
also SC-FM 
bilayers. We
have shown that, because of the numerous parameters for the
system, such as
temperature, magnetization, thickness etc., these systems display a
complex
phase diagram. We did not exhaust all possible states. Moreover, 
dynamic
effects and transport properties are beyond the scope of this
article.
However, we believe that our primary  results are of experimental 
interest
and have technological promise.

We believe that the most experimentally interesting and challenging
predictions are: (1), the existence of domain structures with
alternating
magnetization and vortex polarity in FM-SC bilayers and their
shrinking below the SC transition temperature; (2), the formation of
several vortices  under the magnetic dots placed on top of a SC film;
(3), the symmetry violation in the periodic array of the
magnetic dots on SC films.
Although, antivortices outside the single magnetic dot are not
analyzed
in this article, they are expected to occur in real systems. All 
these
effects can be observed directly by
scanning tunnelling microscopy, scanning Hall probe microscopy and 
micro
SQUID measurements.

In conclusion, our results not only confirm some old results found by
means of different methods, but also present a class of new physical
effects in HMSS. In this sense, they manifest a new direction and
motivation for the possible experiments in the future.

\begin{appendix}
\section{The Pearl Vortex}

The vortices in thin superconducting(SC) films are first studied by J. 
Pearl 
\cite{pearl}. Here, we give the detailed calculations of vector potential 
and magnetic field for Pearl vortex located at  ${\bf r}_0$ on the SC 
film. We 
start with 
the London-Pearl equation:

\begin{equation}
\nabla^2 {\bf A} ( {\bf r}, z ) = \frac{1}{\lambda} {\bf a} ( {\bf r} ) 
\delta ( z ) - \frac{\phi_0}{ 2 \pi \lambda } {\bf \nabla} \varphi ( {\bf 
r} - {\bf r}_0 ) \delta ( z ), 
\label{plon}
\end{equation}

\noindent where ${\bf a} ( {\bf r} ) = {\bf A} ( {\bf r}, z = 0 )$.  
It is easy to find the vector potential due to the vortex by employing the 
Fourier transformation to Eq.(\ref{plon}). In doing so, we use

\begin{equation}  
{\bf A}  ( {\bf r}, z ) = \int {\bf A}_{\bf k} e^{i {\bf q}\cdot{\bf r} + 
i k_z  z} \frac{d{\bf k}}{(2 \pi)^3}.
\label{fdef}
\end{equation}
\noindent  In the Fourier 
representation, the London-Pearl equation reads

\begin{equation}
{\bf A}_{\bf k} = - \frac{1}{\lambda k^2} {\bf a}_{\bf q} + 
\frac{\phi_0}{2 \pi \lambda} \frac{({\bf \nabla} \varphi)_{\bf q}}{k^2},
\label{fourlon}
\end{equation}

\noindent where $({\bf \nabla} \varphi)_{\bf q} = 2 \pi \frac{i \hat q 
\times \hat z}{q} e^{-i {\bf q} \cdot {\bf r}_0}$ \cite{abr}. Employing 
integral over $k_z$ to Eq.(\ref{fourlon}) and using ${\bf a}_{\bf q} = 
\int_{-\infty}^{\infty} \frac{d k_z}{2 \pi} {\bf A}_{\bf k}$, we obtain 
the Fourier-transform of the vortex-induced vector potential at the SC 
film as

\begin{equation}
{\bf a}_{\bf q} = \frac{i \phi_0 ( \hat q \times \hat z ) e^{-i {\bf q} 
\cdot {\bf r}_0}}{q  ( 1 + 2 \lambda q )}.
\label{avq}
\end{equation}

\noindent Substituting the above equation  in Eq.(\ref{fourlon}), the 3d 
vortex-induced vector potential is found as

\begin{equation}
{\bf A}_{\bf k} = \frac{ 2 i \phi_0 (  \hat q \times \hat z ) e^{-i {\bf 
q}\cdot{\bf r}_0}}{{\bf k}^2 ( 1 + 2 \lambda q )}.
\label{3dav}
\end{equation}

\noindent The direct substitution of Eq.(\ref{3dav}) back in 
Eq.(\ref{fdef}) leads to

\begin{equation}
{\bf A} ({\bf r}, z ) = \int \frac{ 2 i \phi_0 (  \hat q 
\times \hat z ) 
e^{-i {\bf
q}\cdot({\bf r}-{\bf r}_0) + i k_z z}}{{\bf k}^2 ( 1 + 2 \lambda q )} 
\frac{d^2q dk_z}{(2 \pi)^3}.
\label{3dv}
\end{equation}

\noindent First, we perform integral over $k_z$,  and find

\begin{equation}
{\bf A} ({\bf r}, z ) = \int \frac{  i \phi_0 (  \hat q 
\times \hat 
z )
e^{-i {\bf
q}\cdot({\bf r}-{\bf r}_0)} e^{-  q|z|}}{q( 1 + 2 \lambda q )}
\frac{d^2q }{(2 \pi)^2}.
\label{3dvnew}
\end{equation}

\noindent Let 

\begin{eqnarray}
{\bf q} &=& q \cos ( \theta + \phi ) \hat u + q \sin  ( \theta + \phi )  
\hat v \label{q} \\
{\bf r} - {\bf r}_0 &=& | {\bf r} - {\bf r}_0 | \cos \phi \hat u + |{\bf 
r} 
- {\bf r}_0| \sin \phi \hat v,
\label{qandr}
\end{eqnarray}
where $\hat u$ and $\hat v$ are unit vectors in a plane perpendicular to 
the $z$-direction,  $\theta$ is the angle between ${\bf q}$ and ${\bf r} - 
{\bf r}_0$, and $\phi$ is the angle between ${\bf r} - {\bf r}_0$ and 
$\hat u$. By (\ref{q}), $\hat q \times \hat z$ in Eq.(\ref{3dvnew}) reads
\begin{equation}
\hat q \times \hat z  = - \cos ( \theta + \phi ) \hat v + \sin  ( \theta + 
\phi ) \hat u.
\label{qtimezz}
\end{equation}
Substituting (\ref{qtimezz}) in Eq.(\ref{3dvnew}), and using $\hat u 
\times \hat z = - \hat v$, $\hat v
\times \hat z = \hat u$,  
$\int_0^{2 \pi} e^{i x \cos \theta} \cos \theta d \theta
= 2 \pi i J_1 ( x )$ and $\int_0^{2 \pi} e^{i x \cos \theta} \sin \theta d 
\theta
= 0$, we find

\begin{equation}
{\bf A}({\bf r},z)
=
\frac{n \phi_0}{2 \pi}
\frac{\hat z\times ({\bf r}-{\bf r}_0) }{\vert {\bf r}-{\bf r}_0  
\vert}
\int_{0}^{\infty} \frac{J_1 ( q\vert{\bf r} - {\bf
r}_0\vert) e^{-q| z |}} {1 + 2\lambda q}dq. \label{vec2}   
\end{equation}.
 
\noindent Note that in the above equation, $\cos\phi \hat v - 
\sin \phi 
\hat u = \frac{\hat 
z\times ({\bf r}-{\bf r}_0) }{\vert {\bf r}-{\bf r}_0
\vert}$. From ${\bf B} = {\bf \nabla} \times {\bf A}$, 
the magnetic field components of the Pearl vortex are found as follows:

\begin{eqnarray}
B_r ({\bf r},z)
&=&
\frac{n \phi_0}{2 \pi}\mbox{sign}(z)
\int_{0}^{\infty} \frac{J_1 ( q\vert{\bf r} - {\bf
r}_0\vert) q e^{-q| z |}} {1 + 2 \lambda q}dq, \\  
B_z ({\bf r},z)
&=& \frac{n \phi_0}{2 \pi}\int_{0}^{\infty} 
\frac{J_0 ( q\vert{\bf r} 
- {\bf
r}_0\vert) q e^{-q| z |}} {1 + 2 \lambda q}dq.
\end{eqnarray}

\section{Integrals of Bessel Functions}

\noindent In this appendix, the asymptotic values  and the exact results 
of the 
Bessel 
integrals used in this article are introduced. First, we present the 
integrals 
containing  one Bessel function. These integrals are in the form of 
\begin{equation} 
\int_{0}^\infty \frac{J_m ( k r ) k^n}{1 + 2 k \lambda} d 
k,
\label{C1}
\end{equation}
where $m=0,1$ and $n=0,1$.  For $m=0,1$ and $n=0$, the exact results can 
be 
obtained as follows \cite{bessel}:

\begin{equation}
\int_{0}^{\infty} \frac{J_0(kr)}{1+2\lambda k} dk = \frac{\pi}{4 
\lambda}[H_0 ( \frac{r}{2 \lambda}) - Y_0 ( \frac{r}{2 \lambda})],
\label{e1}
\end{equation}

\begin{equation}
\int_{0}^{\infty} \frac{J_1(kr)}{1+2 k \lambda} dk = \frac{\pi}{4
\lambda}[H_{-1} ( \frac{r}{2 \lambda}) + Y_1 ( \frac{r}{2 \lambda})]
+\frac{1}{r},
\label{e2}
\end{equation}

\noindent where $H_l ( x )$ and $Y_l ( x)$ are the Struve function and 
the second kind of the Bessel function of the $l$th order \cite{abrom}. 
The respective asymptotic values  of Eqs.(\ref{e1}, \ref{e2}) 
are found as follows: When $r << \lambda$, which corresponds to $k 
\lambda >> 1$, the integral in (\ref{e1}) 
becomes \cite{GR} 
\begin{equation} 
\int_{0}^{\infty} \frac{J_0(kr)}{2 k \lambda} dk
= \frac{1}{2 \lambda}[\ln(\frac{\lambda}{r}) - C].
\end{equation}
For $r>>\lambda$, which is equivalent to $k \lambda  << 1$, 
the integral in (\ref{e1}) becomes \cite{GR}
\begin{equation}
\int_{0}^{\infty} J_0(kr) dk = \frac{1}{r}. 
\end{equation}
Using the same techniques, the asymptotic values of the integral in 
(\ref{e2}) are 

\begin{eqnarray}
\int_{0}^{\infty} \frac{J_1(kr)}{1+2 k \lambda} dk &\approx& 
\int_{0}^{\infty} \frac{J_1(kr)}{2 k \lambda} dk \,=   \frac{1}{2 
\lambda}
\;\;\:\:\;
r << \lambda \\
\int_{0}^{\infty} \frac{J_1(kr)}{1+2 k \lambda} dk &\approx&   
\int_{0}^{\infty} J_1 ( kr) dk \;=\,  \frac{1}{r} 
\;\;\;\;\;\ r 
>> 
\lambda.
\end{eqnarray}

\noindent In order to find the asymptotic values for $m = 0,1$ and $n = 
1$,  we use   

\begin{equation}
\int_0^\infty J_m ( k r ) k^n d k = \mbox{lim}_{\alpha \rightarrow 0}
\frac{\partial^n}{\partial \alpha^n} \int_0^\infty J_m ( k r ) e^{-\alpha
k} d k.
\label{tech}
\end{equation}

\noindent With similar 
techniques and 
(\ref{tech}), the asymptotic values are given as follows:

\begin{equation}
\int_{0}^{\infty} \frac{J_0(kr)}{1+2 k \lambda}k dk 
\approx \int_{0}^{\infty} \frac{J_0(kr)}{2 \lambda}dk \; = 
\frac{1}{2 \lambda r}  \;\;\;\;
r << \lambda. 
\label{j0k1}
\end{equation}

\noindent For $r>>\lambda$, the integral becomes $\int_{0}^{\infty} 
J_0(kr)k dk$,  which equals zero \cite{GR}. In order to find nonzero 
result, we do 
the following approximation: $r>>\lambda$ is equivalent to $k \lambda << 
1$. 
Therefore, the fraction in (\ref{C1}) can be rewritten as  

\begin{equation}
\frac{1}{1 + 2 k \lambda} = 1 - 2 k \lambda + \cdots
\label{app2}
\end{equation}
By (\ref{app2}), we obtain
\begin{equation}
\int_{0}^{\infty} \frac{J_0(kr)}{1+2 k \lambda}k dk \approx 
- 2 \lambda \int_{0}^{\infty} J_0(kr)k^2 dk =  \frac{2 
\lambda}{r^3} \;\;\;\; r >> \lambda,
\label{j0k0}
\end{equation}

\noindent and, 

\begin{eqnarray}
\int_{0}^{\infty} \frac{J_1(kr)}{1+2 k \lambda}k dk &\approx& 
\int_{0}^{\infty} \frac{J_1(kr)}{2 \lambda} dk \; = 
\frac{1}{2 \lambda r}  \;\;\;\;
r << \lambda \\
\int_{0}^{\infty} \frac{J_1(kr)}{1+2 k \lambda}k dk &\approx& 
\int_{0}^{\infty} J_1(kr) k dk = \frac{1}{r^2}  
\;\;\;\;\;\ r >> \lambda.
\label{J1k1}
\end{eqnarray}

\noindent We give only the asymptotic values  of the integrals with two 
Bessel 
functions. We start with the following integral:
\begin{equation}
\int_{0}^{\infty} \frac{J_1(kR)J_0(kr)}{1+2 k \lambda} dk. 
\label{2bc1}
\end{equation}
For the above integral, we first analyze the case, in which  $R<<\lambda$. 
In this case, we can replace  $J_1 ( k R )$ by  $k R/2$ in    
(\ref{2bc1}). In doing so, we get
\begin{equation}
\int_{0}^{\infty} \frac{J_1(kR)J_0(kr)}{1+2 k \lambda} dk \approx 
\frac{R}{2} \int_{0}^{\infty} \frac{J_0(kr)}{1+2 k \lambda} k dk.  
\end{equation}
Using the asymptotic values for the integral on the left in 
(\ref{j0k1}) and (\ref{j0k0}), we find

\begin{eqnarray}
\int_{0}^{\infty} \frac{J_1(kR)J_0(kr)}{1+2 k \lambda} dk &=& 
\frac{ R}{4\lambda r}
\;\;\;\;\;\;\;
R << r << \lambda \\  
&=& \frac{2 
\lambda R}{r^3} \;\;\;\;\;\;  R << \lambda << r. 
\end{eqnarray}
For $R>\lambda$ and $r>\lambda$, we can neglect $2 k \lambda$ in 
Eq.(\ref{2bc1}). In doing so, we obtain 
\begin{equation}
\int_{0}^{\infty} \frac{J_1(kR)J_0(kr)}{1+2 k \lambda} dk \approx
\int_{0}^{\infty} J_1(kR) J_0(kr)dk.
\end{equation}  
The above integral equals \cite{GR}

\begin{eqnarray}
\int_{0}^{\infty} J_1(kR) J_0(kr)dk &=& \;\ 0 
\;\;\;\;\;\;\;\;\;\;\;\;R < r  \\
&=& \frac{1}{2R}\;\;\;\;\;\;\;\;\;\ R = r \\   
&=& \frac{1}{R} \;\;\;\;\;\;\;\;\;\;\;\ r < R. 
\end{eqnarray}
The other integral of interest containing two Bessel functions is  

\begin{equation}
\int_{0}^{\infty} \frac{J_1(kR)J_0(kr)k^2}{1+2 k \lambda} dk. 
\label{2bess}
\end{equation}
For $R<\lambda$, using $J_1 ( k R ) \approx k R / 2$, the integral in 
Eq.(\ref{2bess}) becomes 
\begin{equation}
\frac{R}{2}\int_{0}^{\infty} \frac{J_0(kr)k^3}{1+2 k \lambda} dk.
\label{2bess1} 
\end{equation}
By (\ref{tech}) and (\ref{app2}), the asymptotic values of 
the above 
integral can be calculated as follows: for $r << \lambda$, (\ref{2bess}) 
can be rewritten as
\begin{equation}
\frac{R}{4 \lambda} \int_{0}^{\infty} J_0(kr) k^2 d k = \frac{R}{4 \lambda 
r^3}.
\end{equation}
For  $R << r$, (\ref{2bess1}) becomes 
\begin{equation}
- R\lambda \int_{0}^{\infty} J_0(kr) k^4 d k = - \frac{9 R \lambda}{r^5}. 
\end{equation} 
Now, we can write the asymptotic values of (\ref{2bess}) as

\begin{eqnarray}
\int_{0}^{\infty} \frac{J_1(kR)J_0(kr)k^2}{1+2 k \lambda} dk &\approx& 
\frac{R}{4 \lambda r^3} \;\;\;\;\;\;\ R << r << \lambda \\ 
\int_{0}^{\infty} \frac{J_1(kR)J_0(kr)k^2}{1+2 k \lambda} dk &\approx&    
 - \frac{9 R \lambda}{r^5}. \;\;\;\; R << \lambda << r.
\end{eqnarray}

\section{Calculation of Series}

In this appendix, the detailed analysis of series is given. First, the 
series in the energy calculations of the periodic systems are analyzed; 
second, the detailed calculation of the vortex density is shown. The 
series  we 
encounter in the energy  calculations fall into two categories. In the 
first category, 
we sum over 
one variable. The series in this category are in the form of 
$\sum_{r=1}^{r_{max}} 1/r$. Employing the Euler-Maclaurin summation 
formula \cite{arfken}, 
the 
summation is found with logarithmic accuracy as
\begin{equation}
\sum_{r=1}^{r_{max}} \frac{1}{r} \approx \ln r_{max} + C.
\label{s1}
\end{equation}
\noindent where $C \sim 0.577$ is the Euler-Mascheroni constant. If the 
summation is performed over only odd integers, we can still transform our 
series to (\ref{s1}). Namely,

\begin{eqnarray}
\sum_{r=0}^{r_{max}} \frac{1}{2 r + 1} &\approx& \sum_{r=1}^{2 r_{max} +1} 
\frac{1}{r} - \frac{1}{2} \sum_{r=1}^{r_{max}/2} \frac{1}{r}, \\ 
&\approx& \ln ( 2 r_{max} + 1 ) + C - \ln (\frac{r_{max}}{2}) - 
\frac{C}{2}, \\
&\approx& \frac{1}{2} ( \ln r_{max} + C + 2 \ln 2 ).
\label{s2new}
\end{eqnarray}

\noindent The second category is the double series. In this aspect, we 
first show 
the 
calculation of square domain energy in the continuum approximation. The 
corresponding energy contains the 
series 
\begin{equation}
S = \sum_{r,s=0}^{\infty} \frac{\sqrt{(2 r + 
1)^2 + (2 s + 1)^2}}{(2 r + 1)^2 ( 2 s + 1)^2}.
\label{sqarra}
\end{equation}
Our goal is to calculate 
the logarithmic contribution due to the self-vortex energies and the 
constant as a next approximation. The 
sum in 
(\ref{sqarra}) diverges 
logarithmically in two regions: $r>>s$ and $s>>r$. Keeping this in mind, 
(\ref{sqarra}) can be rewritten as  

\begin{eqnarray}
S &=& \sum_{r,s=-\infty}^{\infty} \frac{\mbox{max}( 2 r + 1, 2 s +
1)}{(2 r + 1)^2 ( 2 s + 1)^2} \nonumber
\\
&+&  \sum_{r,s=-\infty}^{\infty} \frac{\sqrt{(2 r +1)^2 + (2 s + 1)^2} -
\mbox{max}( 2 r + 1, 2 s +
1)}{(2 r + 1)^2 ( 2 s + 1)^2}.
\label{sqser}
\end{eqnarray}
The first term in the above series contributes the logarithmic term 
$S_{log}$, and here it is 

\begin{eqnarray}
S_{log} &=& 2 \sum_{r_{max}} \frac{1}{2 r + 1} \sum_{s=0}^{\infty} 
\frac{1}{(2 s + 1)^2} \\ 
&\approx& \frac{\pi^2}{8} (\ln r_{max} + C + 2 \ln 2).
\label{slog}
\end{eqnarray}
We used the result in Eq.(\ref{s2new}), and $\sum_{s=0}^{\infty}
\frac{1}{(2 s + 1)^2} = \frac{\pi^2}{8}$ \cite{series}. The other term  
 in (\ref{sqser}) contributes the constant $S_{cons}$. For three regions 
$r = s$, $r>s$ and $s>r$,  the second series  in (\ref{sqser}) is   
rewritten as follows:
\begin{equation}
S_{cons} = (\sqrt{2}-2) \sum_{r=0}^{\infty} \frac{1}{(2 r + 1)^3} + 2
\sum_{r=0}^{\infty} \sum_{s=r+1}^{\infty} \frac{\sqrt{(2
r +1)^2 + (2 s + 1)^2} - (2 r + 2 s + 2)}{(2 r + 1)^2 ( 2 s + 1)^2}.
\label{scons} 
\end{equation}  
\noindent Numerical calculation gives $S_{cons} = - 1.19$.

The other double series 
of interest here 
are in the form of 
\begin{equation}
I ( x ) = \sum_{r= -\infty}^{r=\infty} \sum_{s= 
-\infty}^{s=\infty} 
\frac{1}{x^2 r^2 + s^2},
\label{s4new}
\end{equation}
\noindent  where $x$ is an arbitrary constant. Although (\ref{s4new}) is 
logarithmically divergent, the sum over one of the variables can be 
done  easily. To this end, we perform the sum over $s$ first. 
In doing so, Eq.  (\ref{s4new}) 
becomes $(2 \pi/x) \sum_{r=1}^{\infty} \coth (\pi x r)/r$ \cite{series}.
This series  is logarithmically divergent. In order to  get the 
logarithmic term , we add and subtract $1/r$. Using the result in 
(\ref{s1}), finally we get

\begin{equation}
I ( x ) \approx \frac{2 \pi}{x} [\sum_{r =1}^\infty \frac{\coth ( \pi x r) 
- 
1}{r} + \ln r_{max} + C].
\label{s5}
\end{equation}   

\noindent Employing the same techniques, we give the results of the 
different versions of Eq. (\ref{s4new}) below:

\begin{equation} 
\sum_{r= -\infty}^{r=\infty} \sum_{s=
-\infty}^{s=\infty}
\frac{1}{x^2 (2 r + 1)^2 + s^2} \approx
 \frac{2 \pi}{x} [\sum_{r=0}^{\infty} \frac{\coth (( 2 r + 1)\pi x ) -
1}{2 r + 1} + \frac{\ln r_{max}}{2} + \frac{C}{2}], 
\label{b14}
\end{equation}

\begin{equation}
\sum_{r= -\infty}^{r=\infty} \sum_{s=
-\infty}^{s=\infty}
\frac{1}{x^2 (2 r + 1)^2 + ( 2 s+ 1)^2} \approx
 \frac{ \pi}{x} [\sum_{r=0}^{\infty} \frac{\tanh (( 2 r + 1)\frac{\pi 
x}{2} ) -
1}{2 r + 1} + \frac{\ln r_{max}}{2} + \frac{C}{2}]. 
\label{b15}
\end{equation}

\noindent In (\ref{b14}) and (\ref{b15}), we use $\sum_{s=0}^{\infty} 
1/(y^2 + ( 2 s + 1 )^2) = \pi 
\tanh(\pi y/2)/(4 y)$. In the presence of $\sin^2 ( (2 r + 1 ) y)$ or 
$\cos^2 ( (2 r + 1 ) y)$, the series can be calculated in a similar way, 
using $\sin^2 ( (2 r + 1 ) y) = (1 - \cos(2 ( 2 r + 1)y))/2$ or  $\cos^2 ( 
(2 r + 1 ) y) = (1 + \cos(2 ( 2 r + 1)y))/2$. For example, 

\begin{eqnarray}
\sum_{r= -\infty}^{r=\infty} \sum_{s=
-\infty}^{s=\infty} 
\frac{\sin^2(( 2 r + 1) y)}{(x^2 (2 r+ 1)^2 + s^2)}
 &=& \frac{2 \pi}{x} [\sum_{r=0}^{\infty} \frac{\sin^2 ( (2 r + 1 ) y 
)(\coth (( 2 r + 1)\pi x ) -
1)}{2 r + 1}\nonumber \\ 
&+& \frac{\ln r_{max}}{4} - \frac{\ln |\cot(y/2)|}{4}+  \frac{C}{4}],
\end{eqnarray}

\begin{eqnarray}
\sum_{r= -\infty}^{r=\infty} \sum_{s=
-\infty}^{s=\infty}
\frac{\cos^2(( 2 r + 1) y)}{(x^2 (2 r+ 1)^2 + s^2)}
 &=& \frac{2 \pi}{x} [\sum_{r=0}^{\infty} \frac{\sin^2 ( (2 r + 1 ) y 
)(\coth (( 2 r + 1)\pi x ) -
1)}{2 r + 1}\nonumber \\
&+& \frac{\ln r_{max}}{4} + \frac{\ln |\cot(y/2)|}{4}+  \frac{C}{4}].
\end{eqnarray}

\noindent We use 
\begin{equation}
\sum_{r=0}^\infty \frac{\cos((2 r +1 )\theta)}{2 r +1 } = \frac{\ln | 
\cot 
(\theta/2)|}{2}. 
\label{coth}
\end{equation}

The Fourier transform  of the vortex density $n_{\bf G}$ for the stripe 
domain 
structure is found by substituting the corresponding reciprocal lattice 
vectors $G = \pi ( 2 r + 1 )/L$ and the Fourier transform of the step 
function $s_{\bf G} = \frac{2 i}{\pi ( 2 r + 1)}$ in Eq.(\ref{dens-F}). In 
doing so, we obtain
\begin{equation} 
n_{\bf G} =  - \frac{4 \pi i \tilde 
\varepsilon_v}{L \phi_0^2} \sum_{r=-\infty}^{\infty} \mbox{sign} ( 2 r + 
1).
\label{nfour}
\end{equation}

\noindent Employing  the inverse Fourier transform to 
Eq.(\ref{nfour}) gives the following series:

\begin{equation}
F ( x ) =  \sum_{r=-\infty}^{\infty} \mbox{sign} ( 2 r + 1) e^{i ( 2 r + 1 
) 
\frac{\pi x}{L}} = 2 i \sum_{r=0}^{\infty} \sin (( 2 r + 1 )\frac{\pi 
x}{L}).
\end{equation}

\noindent In order to calculate the above series, we integrate both 
sides over $x$. In doing so, by (\ref{coth}) we find

\begin{equation}
\int F( x ) d x = - 2 i \frac{L}{\pi} \sum_{r=0}^{\infty} \frac{\cos ( ( 2 
r + 1 )\frac{\pi
x}{L})}{2 r + 1} = - i \frac{L}{\pi} \ln |\cot ( \frac{\pi x }{2 L})|. 
\end{equation}

\noindent The derivative of the above 
equation with respect to 
$x$ gives

\begin{equation}
F(x) = - \frac{i}{\sin ( \frac{\pi x}{L})}.
\label{sinf}
\end{equation}

\noindent By (\ref{nfour}) and (\ref{sinf}), the vortex density $n(x)$ 
becomes

\begin{equation}
n(x) = - \frac{4 \pi \tilde
\varepsilon_v}{L \phi_0^2} \frac{1}{\sin ( \frac{\pi x}{L})}.
\end{equation}

\end{appendix}

\pagebreak

\end{document}